\documentclass[sigconf]{acmart}

\usepackage{graphicx}
\usepackage{balance}  % for  \balance command ON LAST PAGE  (only there!)

\usepackage{caption}
\usepackage{subcaption}
\usepackage{soul}
\usepackage[linesnumbered,ruled]{algorithm2e}
\usepackage{algorithmic}
\usepackage{ifthen}
\usepackage{amsfonts}

\usepackage{booktabs} % For formal tables
\usepackage{enumitem}
\usepackage{comment}

\usepackage{hyperref}

\usepackage{color, colortbl}
\definecolor{Gray}{gray}{0.9}

\usepackage{multirow}

\usepackage{relsize}

\usepackage{setspace}
\usepackage{microtype}

\copyrightyear{2019} 
\acmYear{2019} 
\setcopyright{acmcopyright}
\acmConference[KDD '19]{The 25th ACM SIGKDD Conference on Knowledge Discovery and Data Mining}{August 4--8, 2019}{Anchorage, AK, USA}
\acmBooktitle{The 25th ACM SIGKDD Conference on Knowledge Discovery and Data Mining (KDD '19), August 4--8, 2019, Anchorage, AK, USA}
\acmPrice{15.00}
\acmDOI{10.1145/3292500.3330755}
\acmISBN{978-1-4503-6201-6/19/08}

\begin{document}

\title[Applied Data Science Track Paper]{Short and Long-term Pattern Discovery Over Large-Scale Geo-Spatiotemporal Data}

\author[S. Moosavi et al.]{Sobhan Moosavi, Mohammad Hossein Samavatian, Arnab Nandi, Srinivasan Parthasarathy, and Rajiv Ramnath}
\affiliation{%
  \institution{Department of Computer Science and Engineering}
  \institution{The Ohio State University}
  \streetaddress{395 Dreese Laboratories, 2015 Neil Avenue}
  \city{Columbus, Ohio 43210-1277}
}
\email{{moosavi.3,samavatian.1,nandi.9,parthasarathy.2,ramnath.6}@osu.edu}

\begin{abstract}
Pattern discovery in geo-spatiotemporal data (such as traffic and weather data) is about finding patterns of collocation, co-occurrence, cascading, or cause and effect between geospatial entities. Using simplistic definitions of spatiotemporal neighborhood (a common characteristic of the existing general-purpose frameworks) is not semantically representative of geo-spatiotemporal data. 
We therefore introduce a new geo-spatiotemporal pattern discovery framework which defines a semantically correct definition of neighborhood; and then provides two capabilities, one to explore {\em propagation} patterns and the other to explore {\em influential} patterns. 
Propagation patterns reveal common cascading forms of geospatial entities in a region. Influential patterns demonstrate the impact of temporally long-term geospatial entities on their neighborhood.  
We apply this framework on a large dataset of {\em traffic and weather} data at countrywide scale, collected for the contiguous United States over two years. Our important findings include the identification of 90 common propagation patterns of traffic and weather entities (e.g., $rain \rightarrow accident \rightarrow congestion$), which results in identification of four categories of states within the US; and interesting influential patterns with respect to the ``location'', ``duration'', and ``type'' of long-term entities (e.g., $\textit{a major construction} \rightarrow \textit{more traffic incidents}$). 
These patterns and the categorization of the states provide useful insights on the driving habits and infrastructure characteristics of different regions in the US, and could be of significant value for applications such as urban planning and personalized insurance. %Our dataset is also available at \url{https://smoosavi.org/datasets/lstw}. 
\end{abstract}

\begin{CCSXML}
<ccs2012>
<concept>
<concept_id>10010405.10010481.10010485</concept_id>
<concept_desc>Applied computing~Transportation</concept_desc>
<concept_significance>500</concept_significance>
</concept>
<concept_id>10002951.10003260.10003277.10003281</concept_id>
<concept_desc>Information systems~Traffic analysis</concept_desc>
<concept_significance>500</concept_significance>
</concept>
<concept>
<concept_id>10002951.10002952.10003219</concept_id>
<concept_desc>Information systems~Information integration</concept_desc>
<concept_significance>300</concept_significance>
</concept>
<concept>
<concept_id>10002951.10002952.10003219.10003218</concept_id>
<concept_desc>Information systems~Data cleaning</concept_desc>
<concept_significance>300</concept_significance>
</concept>
<concept>
</ccs2012>
\end{CCSXML}

\ccsdesc[500]{Applied computing~Transportation}
\ccsdesc[500]{Information systems~Traffic analysis}
\ccsdesc[300]{Information systems~Information integration}
\ccsdesc[300]{Information systems~Data cleaning}

\keywords{Propagation Patterns, Influential Patterns, Geo-Spatiotemporal Data}

\maketitle

\section{Introduction}
\label{sec:intro}
Spatiotemporal pattern discovery has seen considerable interest over the past decade, with various frameworks were proposed to process the data to find interesting patterns \cite{celik2008mixed,huang2004discovering,yang2005generalized,qian2009mining,mohan2010cascading,mohan2012cascading,liu2011discovering,celik2015partial,yu2016spatial,wajid2016crimestand}. The application domains of relevance include public safety, transportation, earth science, epidemiology, climatology, and environmental management \cite{shekhar2015spatiotemporal}. These frameworks can be used to discover patterns of collocation and co-occurrence, interactions and correlations, cascading, sequential, or cause and effect relationship patterns. However, they all rely on a simplistic definition of spatiotemporal neighborhood, essentially spatial closeness based on an Euclidean or Cartesian system and temporal overlap \cite{huang2004discovering,yang2005generalized,celik2008mixed,mohan2012cascading}, which often makes their use impractical for applications such as traffic, transportation, or weather analyses. For example, a traffic accident on one lane of a freeway has no impact on traffic flow on an opposite lane, yet general-purpose frameworks will locate both lanes in a single neighborhood. Another example arises when studying the impact of a snow event (on traffic flow) which continues well past when the snow event has ended. The time overlap constraint required by existing frameworks would hinder such a study. Note that there may not be any trivial changes to be made to make the existing frameworks semantically applicable for this type of data. Because, their basis is on a specific way of defining spatiotemporal neighborhood, which changing that would make them unusable (e.g., regarding their pruning step) or expensive to be employed. 

To address these challenges, we propose a new framework for finding patterns in geo-spatiotemporal data. This framework consists of two parts, one to explore {\em propagation} patterns, and the other to reveal {\em influential} patterns. Identifying propagation patterns requires the exploration of partially ordered sets of geospatial entities, that are spatially co-located and temporally co-occurring, with potential ``cause and effect'' relationships between the entities. An example of this type is a {\tt rain} event, which causes an {\tt accident}, with the accident then causing {\tt congestion}. Identifying influential patterns, on the other hand, requires studying the impact of temporally long-term geospatial entities (e.g. a major construction) on their spatial neighborhoods. An example of this type of pattern is the increase in number of {\tt congestion} events in a region because of a long-term {\tt snowing} event. 

To explore propagation patterns -- also referred as ``cascading patterns''\cite{mohan2012cascading} or ``spatiotemporal couplings'' \cite{shekhar2015spatiotemporal}, we propose a tree-pattern-mining-based process, we term {\em short-term pattern discovery}, which employs a strict definition of spatial neighborhood to ensure spatial collocation, and a definition of temporal co-occurrence specific to geo-spatiotemporal data and application domain constraints. To explore influential patterns -- also referred as ``tele-couplings'' \cite{shekhar2015spatiotemporal} -- we propose a new process, we term {\em long-term pattern discovery}, to examine the effect of long-term entities on their neighborhood to reveal any significant impact. 
As in, and drawing from \cite{hambly2013projected,jaroszweski2014influence}, this process may be used to study impacts with respect to different {\em types}, different {\em locations}, and {\em duration} of long-term geospatial entities. 

To evaluate our framework, we used a large-scale, real-world geo-spatiotemporal dataset of traffic and weather data. This dataset covers the contiguous United States\footnote{The contiguous United States excludes Alaska and Hawaii, and considers District of Columbia (DC) as a separate state.}, includes data collected from August 2016 to August 2018, and contains about $13.1$ million instances of traffic entities (e.g., accident, congestion, and construction), and about $2.2$ million instances of weather entities (e.g., rain, snow, and storm). Through the processes mentioned above, we found 90 common patterns of propagation of relatively short-term traffic or weather entities, and identified {\em four} categories of states based on these patterns. In addition, we carefully studied the impact of relatively long-term traffic or weather entities on traffic, and identified a variety of insights with respect to ``location'', ``type'', and ``duration'' of the entities. The main contributions of this paper are as follows: 
\begin{itemize}[leftmargin=5pt]
  \item Short-term pattern discovery: We propose a new process for discovering propagation patterns in geo-spatiotemporal data, which models spatiotemporal collocation and co-occurrence in terms of tree structures, and adopts an existing tree pattern mining approach to reveal prevalent patterns. In comparison to the general purpose frameworks, this method better suits application domain requirements of a stricter definition of spatiotemporal neighborhood. 
  \item Long-term pattern discovery: We propose a new process for discovering influential patterns in geo-spatiotemporal data, which examines the impact of long-term geospatial entities on their neighborhood in order to reveal significant influential patterns. Exploring such patterns with existing frameworks is not feasible, due to lack of effective spatiotemporal neighborhood metrics to explore longer-term (or lagging) impacts. 
  \item Data collection and processing: We present a set of processes for collecting real-time traffic and historic weather data, using which we built a publicly available ``research dataset'' of $13.1$ million traffic entities (e.g., accident, congestion, and construction), and $2.2$ million weather entities (e.g., rain, snow, and storm). This dataset is accessible from \url{https://smoosavi.org/datasets/lstw}. 
  \item Findings and insights: By applying our new framework on the above dataset, we present a range of insights for different regions in the United States. These insights may be further utilized for applications such as urban planning, exploring flaws in transportation infrastructure design, traffic control and prediction, impact prediction, personalized insurance, potentially with relevance to the creation of smart cities. 
\end{itemize}

The rest of this paper is organized as follows: We review the related work in Section~\ref{sec:rel}, and provide preliminaries in Section~\ref{sec:prob}. Section~\ref{sec:data} describes the dataset preparation, followed by description of framework in Section~\ref{sec:method}. Experiments and results are presented in section~\ref{sec:results}, and Section~\ref{sec:conc} concludes the paper. 

\vspace{-5pt}
\section{Related Work}
\label{sec:rel}
Spatiotemporal pattern discovery has been thoroughly discussed in literature  \cite{huang2004discovering,celik2008mixed,qian2009mining,mohan2010cascading,liu2011discovering,mohan2012cascading,celik2015partial}. Earlier work focused more on spatial prevalence and paid less attention to temporal aspects \cite{huang2004discovering}, while later work considered both aspects simultaneously \cite{shekhar2015spatiotemporal}. The common process of spatiotemporal pattern discovery is to first define spatiotemporal co-occurrence and collocation criteria; then introduce an interest measure (e.g., participation index); and finally outline a {\em miner} algorithm to find interesting patterns \cite{huang2004discovering}. 
Techniques in these papers being general purpose solutions, rely on simplistic definitions of collocation (spatial) and co-occurrence (temporal), and unable to reveal complex spatiotemporal correlations (such as influential patterns). Further, they have been developed and only tested on small-scale (real-world or synthetic) data. 
To address these challenges with respect to geo-spatiotemporal data, we propose a new framework which provides an appropriate and precise definition of collocation and co-occurrence criteria. Moreover, we outline the process of finding complex spatiotemporal patterns and prove its applicability through extensive experiments. Lastly, we apply our framework on a large-scale, countrywide geo-spatiotemporal dataset of traffic and weather data to explore interesting patterns. 

Regarding the application domain, there are numerous studies for finding patterns in traffic and weather data, with the following goals: to study the impact of precipitation on likelihood or severity of accidents \cite{eisenberg2004mixed,jaroszweski2014influence,theofilatos2017incorporating}; to explore the impact of weather on traffic intensity \cite{cools2010assessing,wang2015predicting}; to reveal the effect of climate change and weather condition on road safety \cite{andersson2011impact,hambly2013projected,theofilatos2014review}; to characterize road accidents locations \cite{kumar2016data}; or, to discover frequent spatiotemporal patterns in traffic data \cite{liu2011discovering,jindal2013spatiotemporal,inoue2016mining}. 
The scale of data in most of these studies is limited to one or at most a few cities. Moreover, interactions and correlations between the different types of traffic entities (accident, congestion, etc.) has not been studied before. Although similar ideas to explore long-term patterns have been previously suggested \cite{eisenberg2004mixed,hambly2013projected,jaroszweski2014influence}, we extend them by: 1) examining a wider range of weather and traffic entity types besides precipitation; 2) exploring properties of different ``locations''; and 3) analyzing the impact of ``duration length'' on traffic flow. 
\vspace{-5pt}
\section{Preliminaries and Problem}
\label{sec:prob}
In this section, we first provide preliminaries and definitions, and then present the problem statement. Note that some of the definitions are customized for our illustration application domain (i.e., traffic and weather data). However, this will not limit their generalizability to the other related domains. %\footnote{Several of definitions in this section are adopted from \cite{tatikonda2006trips}.}. 

\subsection{Definitions}
\begin{itemize}[leftmargin=*]
    \item Geospatial Entity: a geospatial entity $e$ is represented by a tuple $\langle type, start, end, loc\rangle$, which shows an entity of type $type$, happened in time interval $\big[start$, $end\big]$, and its location is specified by $loc$. Definition of $loc$ is related to the application domain. For traffic data, we have $loc = \langle latitude, longitude, Street\_Name,$ $Street\_Side, Zipcode,$ $City,$ $State \rangle$, where $Street\_Side$ shows the relative side of a street (i.e., $R$ or $L$). For weather data, we have $loc = \langle airport\_code\rangle$, which represents the ``airport'' that $e$ is reported from its weather station. A geospatial entity is called {\em long}, if it takes place over a relatively long time interval (see Section~\ref{sec:long}). %Note that we use ``geospatial'' and ``ge-spatiotemporal''  interchangeably throughout the paper. 
    
    \item Weak-Dependency Relationship: two {\em co-occurring} and {\em co-located} geospatial entities are called weakly dependent. Co-occurrence for two entities $e_1$ and $e_2$ means $0 \leq \big|e_1.start - e_2.start\big| \leq \text{\em T-thresh}$, where {\em T-thresh} is a time-threshold. Collocation for two traffic entities requires {\em location matching} as well as {\em spatial closeness}. The former means that all location fields except the GPS coordinates should be the same. By latter, we require that $dist(e_1, e_2) \leq \text{\em D-thresh}$, where $dist$ is the Haversine distance function \cite{haversine} based on GPS coordinates, and {\em D-Thresh} is a distance threshold. With respect to matching a pair of weather and traffic entities, collocation means a match between the ``airport station'' at which the weather entity is reported and the ``airport station'' closest to the traffic entity's location. 
    
    \item Child-Parent Relationship: for two weakly dependent geospatial entities $e_1$ and $e_2$, $e_1$ is a parent for $e_2$ if $e_1$ begins before $e_2$. We treat parent-child relationship as indicative of a {\em cause and effect} relation. A weather entity may only be the parent (or cause) of a traffic entity, and we do not define such a relationship between two weather entities. 
    
    \item Tree Structure: given a set of vertices $\mathcal{V} = \big( v_1, v_2, \dots, v_n\big)$, we define tree $T = (V, E)$, where $V \subset \mathcal{V}$ and $E = \{e_1, e_2,$ $\dots, e_m\}$ is a set of edges, and each edge $e \in E$ connects a pair of vertices $v_i,v_j \in V$ using an {\em un-directed} edge. A tree is an {\em acyclic} graph, and vertices with the same parent are {\em siblings}. Trees in this work have a {\em root} node, sibling nodes are {\em un-ordered}, and nodes are {\em labeled}. Figure~\ref{fig:forest}-(a) shows several examples of such tree structure. In this work, each node of a tree is a geospatial entity, and each edge shows a child-parent relationship between two entities. 
    
    \item Embedded Subtree: given a tree $T=(V,E)$, we define a subtree as $S=(V',E')$, where $V'\subset V$ and $E' \subset E$. A subtree $S$ is said to be an {\em embedded subtree} of $T$ if for each edge $e = (v_a, v_b) \in E'$, $v_a$ is an ancestor (and not necessarily the parent) of $v_b$ in $T$. 
\end{itemize}

\subsection{Short and Long-term Pattern Discovery}
We now formalize the two related problems studied in this paper. 

\subsubsection{Short-term Pattern Discovery}
\begin{figure}
    \centering
    \includegraphics[scale=0.45]{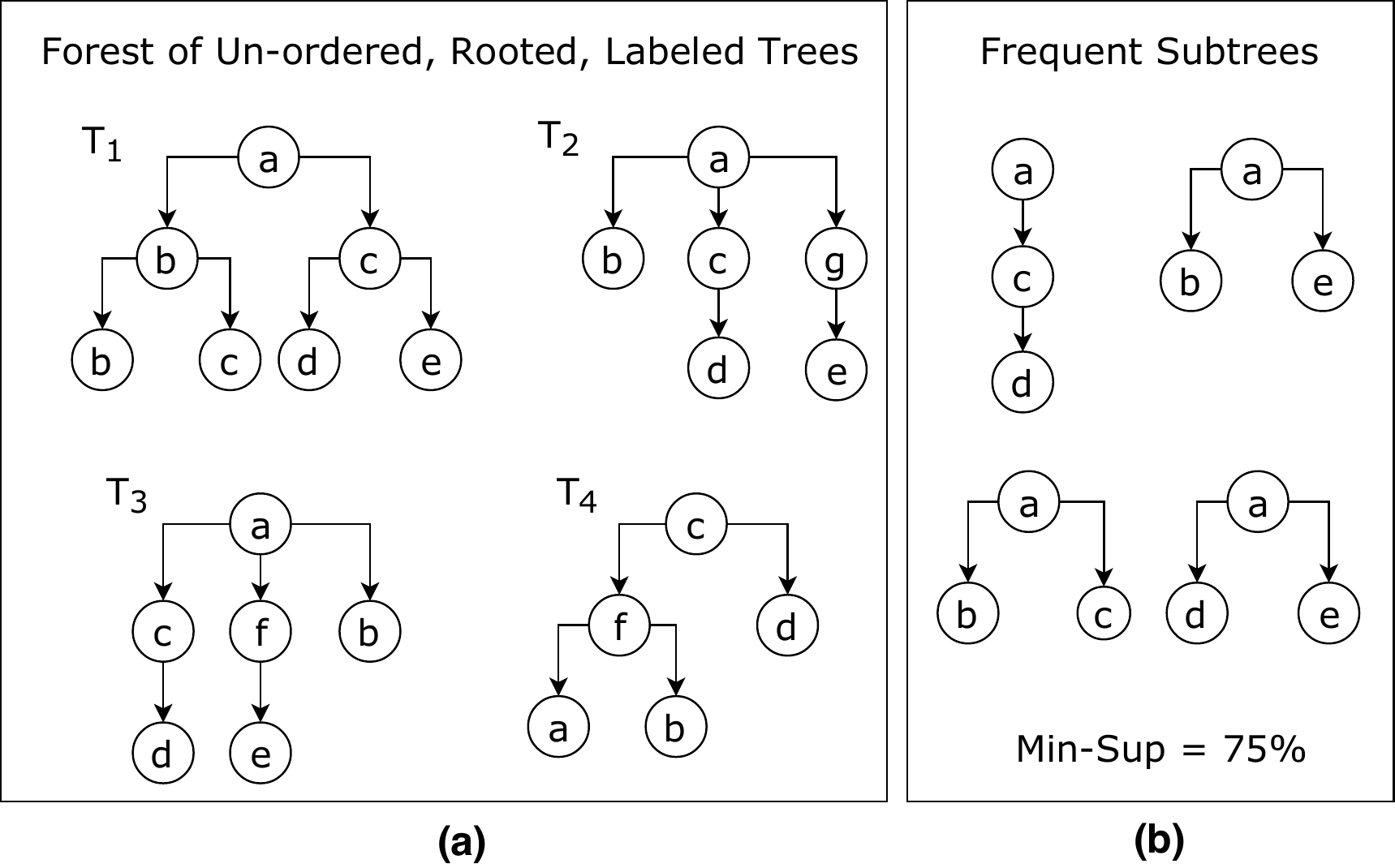}\vspace{-5pt}
    \caption{(a) A forest of four trees, (b) Four of embedded frequently occurred subtrees with a minimum support $75\%$.}
    \label{fig:forest}
    \vspace{-10pt}
\end{figure}

Here we seek to find common short-term {\em propagation patterns} that indicate {\em how} geospatial entities cause other entities to happen. We represent a set of weakly dependent geospatial entities as un-ordered, rooted, labeled trees, where the entities are nodes, weak dependency relations are the edges, and entity types (e.g., rain, accident, and congestion) are the labels of the nodes. Thus, given a forest $F = \{T_1, T_2, \dots, T_k\}$ of such trees, the short-term pattern discovery problem is about finding all embedded subtrees in $F$ which are occurred relatively frequently. Formally, for a subtree $S$ and tree $T$ we define $support(S,T)$ by Equation~~\ref{eq:sup}:
\begin{equation}
    \label{eq:sup}
    support(S, T)=\left\{
            \begin{array}{ll}
              \textrm{1   } \textrm{  if   } S \textrm{   is a subtree of    } T\\
              \textrm{0  } \textrm{  otherwise}
            \end{array}
          \right.
\end{equation}
Then, we define $support(S,F)$ by Equation~\ref{eq:support}: 
\begin{equation}
    \label{eq:support}
    support(S,F) = \frac{\sum_{T \in F} support(S,T)}{|F|}
\end{equation}
For a subtree $S$, if $support(S,F) \geq min\_sup$, where $min\_sup$ is a minimum support threshold, then we say $S$ is a frequent embedded subtree in $F$. An example of a forest with some of frequently occurring subtree patterns is shown in Figure~\ref{fig:forest}. In this example, we have a forest which includes four trees. Using a minimum support threshold of $75\%$, we identified several frequently occurring embedded sub-tree patterns, four of which are shown in Figure~\ref{fig:forest}-(b). 
We use ``short-term pattern discovery'' to indicate that we search for patterns of immediate or short-term impacts, as opposed to long-term impacts which is discussed next. 

\subsubsection{Long-term Pattern Discovery}
Long-term pattern discovery is about exploring the {\em magnitude of impact} of long-term geospatial entities on their neighborhood. As an example, consider a {\em major construction} event in region $A$, because of which, we might observe more congestion events in the same region (when compared to a time when there was not such a construction event). Given a long entity $L$, let $\mathcal{S}_R = [e_1, e_2, \dots]$ be the set of geospatial entities in the {\em vicinity} of that, where \textit{R} is the maximum distance threshold\footnote{For each $e_i \in \mathcal{S}_R$, its location is within distance $R$ from $L$.}. Let $L.start < e.start$ and $L.end > e.end,\text{  }\forall e \in \mathcal{S}_R$. To study the impact of a long entity, we also define two other sets, $\mathcal{S}\textit{--before}_R$ and $\mathcal{S}\textit{--after}_R$. The former contains all geospatial entities which happened within distance $R$ from $L$, during a time interval of the same length as $L$, but before $L$ started. The latter contains all entities in the same neighborhood as $L$, during a time interval of the same length as $L$, but which happened after $L$ ended. 
Given sets $\mathcal{S}_R$, $\mathcal{S}\textit{--before}_R$, and $\mathcal{S}\textit{--after}_R$, we define the problem of ``long-term pattern discovery'' as exploring any significant difference between size of set $\mathcal{S}_R$ and the other two sets. 
In other words, a statistically significant difference between the number of entities when a long entity like $L$ is present, and the number of entities before or after $L$, shows the magnitude of the impact. We call such an occurrence a long-term or influential pattern. 

\subsubsection{Connection Between Problems}
Short-term pattern discovery is about finding {\em immediate} impacts, and long-term pattern discovery is about exploring the ``long-lastingness'' of impacts (i.e., {\em lagging} impacts). Hence, these two are {\em complementary} problems, with each one focused on a separate aspect of dependency and pattern discovery, while using the same set of input data. 

\begin{figure*}[t]
    \vspace{-5pt}
    \small
    \centering
    \hspace{-25pt}
    \begin{subfigure}[b]{0.38\textwidth}
            \includegraphics[width=\linewidth]{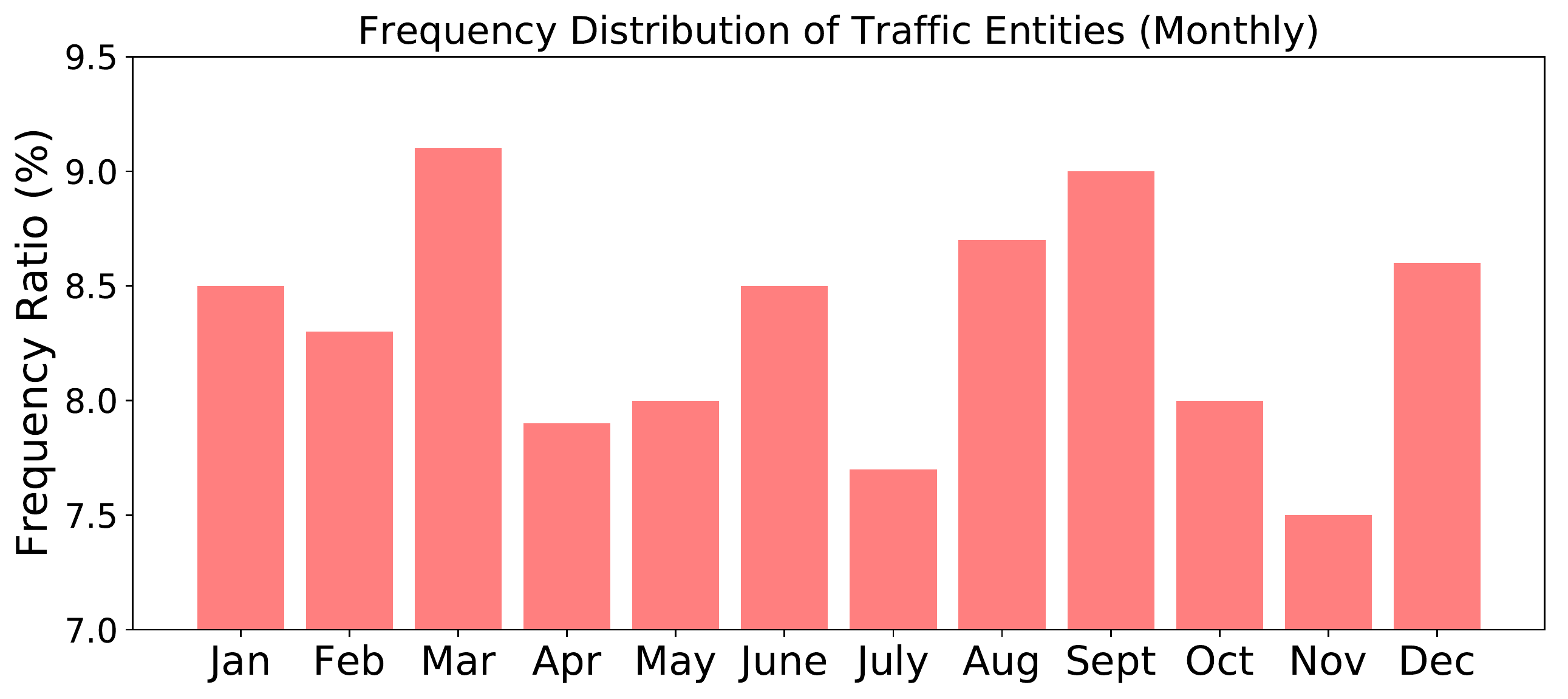}
            \caption{\small Monthly Traffic Distribution}
            \label{fig:traffic_inc}
    \end{subfigure}\hspace{-5pt}
    \begin{subfigure}[b]{0.38\textwidth}
            \includegraphics[width=\linewidth]{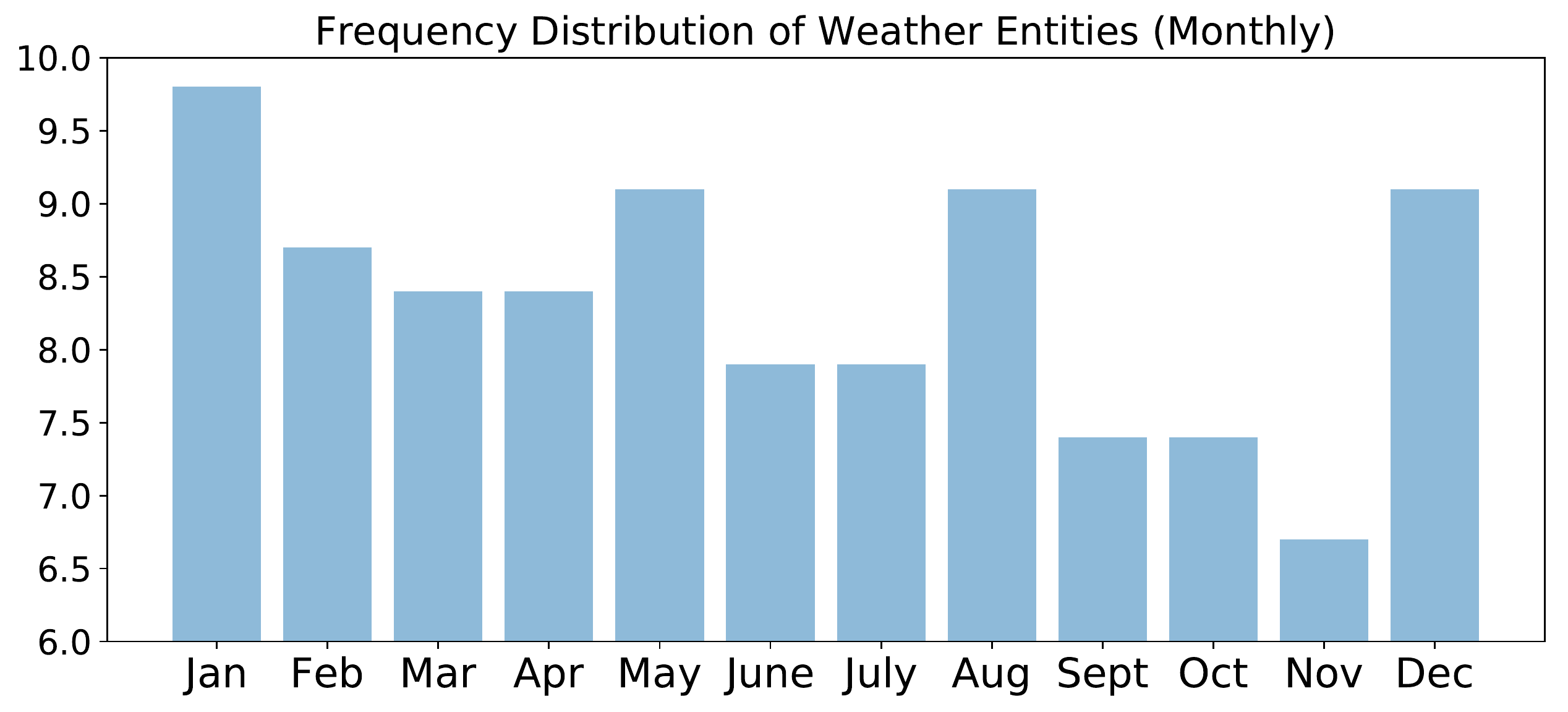}
            \caption{\small Monthly Weather Distribution}
            \label{fig:weather_evn}
    \end{subfigure}\hspace{-5pt}
    \begin{subfigure}[b]{0.287\textwidth}
            \includegraphics[width=\linewidth]{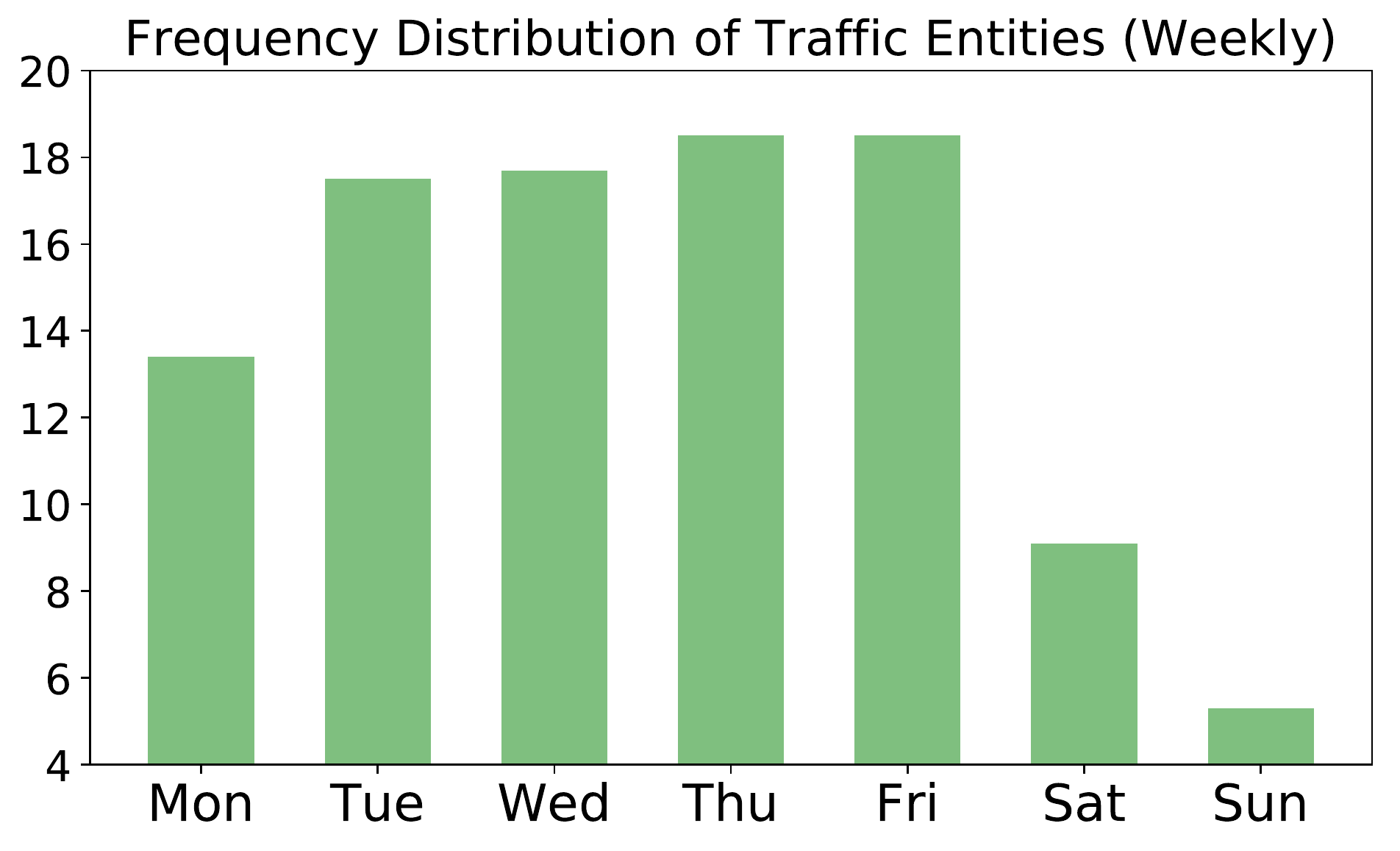}
            \caption{\small Weekly Traffic Distribution}
            \label{fig:traffic_weekly}
    \end{subfigure}
    \hspace{-22pt}
    \vspace{-10pt}
    \caption{\small Relative frequency distribution of traffic and weather data, collected from Aug 2016 to Aug 2018, for the contiguous United States.}
    \label{fig:events_incidents_distribution}
    \vspace{-5pt}
\end{figure*}

\section{Dataset}
\label{sec:data}
In this section, we describe the dataset preparation process. The resulting dataset includes $13.1$ million traffic and $2.2$ million weather entities, which are collected from August 2016 to August 2018. The dataset is available at \url{https://smoosavi.org/datasets/lstw}.

\subsection{Traffic Data}
\subsubsection{Data Collection Process}
\noindent To begin with, traffic entities were collected in real-time using a rest API provided by {\em MapQuest} \cite{mapquest} for a period of two years, from August 2016 to August 2018. To our knowledge, this API broadcast traffic entities captured by a variety of mechanisms - the US and state departments of transportation, law enforcement agencies, traffic cameras, and traffic sensors within the road-networks.  
%provides the same traffic information which is distributed over the Traffic Message Channel (TMC) \cite{ely1990rds}. 
Traffic data was collected for the contiguous United States (49 States). %We pulled data every 90 seconds from 6am to 11pm, and every 150 seconds from 11pm to 6am. 
%by 92 threads, each of which collected data for a bounding box (or geofence) of size $360km \times 360km$ on map. The frequency of API request was every 90 seconds from 6am to 11pm, and every 150 seconds from 11pm to 6am. %To handle the timezone differences, each bounding box was mapped to one of the following timezones: Eastern, Central, Mountain, or Pacific. 
As the raw traffic entities came with GPS coordinates, we employed {\em Nominatim} tool \cite{nominatim} to perform reverse geocoding and translated GPS coordinates to addresses. % which includes attributes such as street, city, state, and zipcode. 

\subsubsection{Data Cleaning Process}
\noindent Following cleaning steps are employed: 
\begin{itemize}[leftmargin=*]
    \item [i.] Resolving duplicates: Duplicates were identified either explicitly by id (i.e., two entities have the same id), or implicitly by content (i.e., two entities of the same type occurring at the same time and location). We kept one entity and removed the other.  
    \item [ii.] Denoising the data: In this context, noise is related to the ``type'' of entity, where the Traffic Message Channel (TMC) \cite{ely1990rds} code (as part of the information for each traffic entity) was different from the default type reported by the MapQuest API. In order to deal with this mismatch, we first extracted 250 different TMC codes from our data, and manually created a new taxonomy by defining a {\em unified type} for each TMC code using \cite{ely1990rds} as reference. Finally, we replaced the new taxonomy with the default one in traffic data.
\end{itemize}

\subsubsection{Data Entity Description}
\noindent We defined the following taxonomy for traffic entities:
\begin{itemize}[leftmargin=10pt]
    \item Accident: a common type, which may involve one or more vehicles, and could result in fatality.  
    \item Broken-Vehicle: refers to the situation when there is one (or more) disabled vehicle(s) in a road. 
    \item Congestion: refers to the situation when the speed of traffic is slower than the expected speed. Using the TMC codes, we defined severity of a congestion as {\em light}, {\em moderate}, or {\em heavy}. 
    \item Construction: an on-going construction or maintenance project on a road.  
    \item Event: situations such as {\em sports event}, {\em concerts}, or {\em demonstrations}, that could potentially impact traffic flow. 
    \item Lane-blocked: refers to the cases when we have blocked lane(s) due to traffic or weather condition.
    \item Flow-incident: refers to all other types of traffic entities. Examples are {\em broken traffic light} and {\em animal in the road}. 
\end{itemize}
Table~\ref{tab:incidents} provides more details on the traffic dataset. The most frequent entity type is ``congestion'' which includes about $80\%$ of the data, and ``accident'' is the second most frequent entity type. Figure~\ref{fig:traffic_inc} also depicts the monthly frequency distribution, where the most entities are observed in March and September and the least in November. Additionally, {\em weekly} frequency distribution of traffic entities is shown by Figure~\ref{fig:traffic_weekly}, where `Friday'' and ``Sunday`` are found to be the days with the most and the least number of recorded entities, respectively.

\begin{table}
    \vspace{-5pt}
    \small
    \centering
    \caption{\small Details on Traffic Dataset, collected for the contiguous United States from Aug 2016 to Aug 2018.}\vspace{-5pt}
    \setlength\tabcolsep{12pt}
    \begin{tabular}{| c | c | c |}
        \hline
        \rowcolor{Gray}
        \textbf{Entity Type} & \textbf{Raw Count} & \textbf{Relative Frequency} \\
        \hline
        Accident & 1,169,507 & 8.9\% \\
        \hline
        Broken-Vehicle & 308,112 & 2.34\% \\
        \hline
        Congestion & 10,542,020 & 80.18\% \\
        \hline
        Construction & 209,933 & 1.60\% \\
        \hline
        Event & 32,817 & 0.25\% \\
        \hline
        Lane-Blocked & 246,832 & 1.88\% \\
        \hline
        Flow-Incident & 637,489 & 4.85\% \\
        \hline
        \hline
        Total & 13,146,710 & 100\% \\
        \hline
    \end{tabular}
    \label{tab:incidents}
    \vspace{-10pt}
\end{table}

\subsection{Weather Data}
\subsubsection{Data Collection Process}
\noindent Raw weather data was collected from 1,973 weather stations located in airports all around the country. The raw data comes in the form of observation records, where each record consists of several attributes such as {\em temperature}, {\em humidity}, {\em wind speed}, {\em pressure}, {\em precipitation} (in millimeters), and {\em condition}\footnote{Possible values are {\em clear}, {\em snow}, {\em rain}, {\em fog}, {\em hail}, and {\em thunderstorm}.}. For each weather station, we receive several observation records per day, which are recorded upon any significant change in any of the measured attributes. 

\subsubsection{Threshold Definition Process}
\noindent To define the taxonomy of weather entities, we require to extract some threshold values. To do so, we used the United State observations of temperature, wind speed, and precipitation amount for rain and snow for a period of {\em seven} years, from January 2010 to January 2016, and applied K-Means clustering algorithm \cite{han2011data} on each of these attributes. The obtained cluster centers are used as threshold for these attributes. 
For temperature, we identified five cluster center values (degrees are in Celsius): $-23.7^\circ$, $-8.6^\circ$, $6.7^\circ$, $21.3^\circ$, and $35.8^\circ$; which we refer them as {\em severe-cold}, {\em cold}, {\em cool}, {\em warm}, and {\em hot}, respectively. For wind speed, we found three cluster centers $13.2 kmh$, $36.2 kmh$, and $60 kmh$, which we refer them as {\em calm}, {\em moderate}, and {\em storm} windy conditions, respectively. For rain, we identified three cluster centers $2.5$, $7.1$, and $11.6$ millimeters, which we refer them as {\em light}, {\em moderate}, and {\em heavy} rainy conditions, respectively. Lastly, for snow we found three cluster centers $0.6$, $1.7$, and $2.5$ millimeters, which we refer them as {\em light}, {\em moderate}, and {\em heavy} snowy conditions, respectively. 

\subsubsection{Entity Extraction Process}
\noindent Given the above threshold values and the raw weather data records from August 2016 to August 2018, we processed each record to use it (if it represents an entity), merge it (if it is part of a previously found entity), or remove it (if it does not represent any entity), and defined the following taxonomy: 
\begin{itemize}[leftmargin=10pt]
    \item Severe-Cold: extremely cold condition, with $temperature \leq -23.7^\circ$.
    \item Fog: low visibility condition as a result of {\em fog} or {\em haze}. 
    \item Hail: solid precipitation including {\em ice pellets} and {\em hail}.
    \item Rain: rain of any type, ranging from {\em light} to {\em heavy}.
    \item Snow: snow of any type, ranging from {\em light} to {\em heavy}.
    \item Storm: the extremely windy condition, where the wind speed is at least $60 kmh$.
    \item Precipitation: any kind of solid or liquid deposit, but different from snow or rain. This was a generic label we frequently observed in raw weather data. %, however, we have no further information to include them in any of the previously described entity types. 
\end{itemize}
We extracted 2,178,949 weather entities for a period of two years. Table~\ref{tab:events} provides more details on weather data, where the most frequent entity types are ``rain'', ``fog'', and ``snow''. Figure~\ref{fig:weather_evn} also shows the frequency distribution of weather entities by month; note that most of the entities occurred in January and the least in November. 
%The source code for weather event extraction based on raw weather data is also available online in our GitHub repository\footnote{See \url{https://github.com/sobhan-moosaviosu/TrafficResearch.}}. 

\begin{table}[h]
    \small
    \centering
    \vspace{-3pt}
    \caption{\small Details on Weather Dataset, collected for the contiguous United States from Aug 2016 to Aug 2018.}\vspace{-8pt}
    \setlength\tabcolsep{12pt}
    \begin{tabular}{| c | c | c |}
        \hline
        \rowcolor{Gray}
        \textbf{Entity Type} & \textbf{Raw Count} & \textbf{Relative Frequency} \\
        \hline
        Severe-Cold & 67,285 & 3.09\% \\
        \hline
        Fog & 454,704 & 20.87\% \\
        \hline
        Hail & 1,252 & 0.06\% \\
        \hline
        Rain & 1,384,588 & 63.54\% \\
        \hline
        Snow & 236,546 & 10.86\% \\
        \hline
        Storm & 14,863 & 0.68\% \\
        \hline
        Precipitation & 19,711 & 0.9\% \\
        \hline
        \hline
        Total & 2,178,949 & 100\% \\
        \hline
    \end{tabular}
    \label{tab:events}
    \vspace{-10pt}
\end{table}

\begin{figure*}[t]
    \small
    \centering
    \hspace{-15pt}
    \begin{subfigure}[b]{0.33\textwidth}
        \includegraphics[scale=0.46]{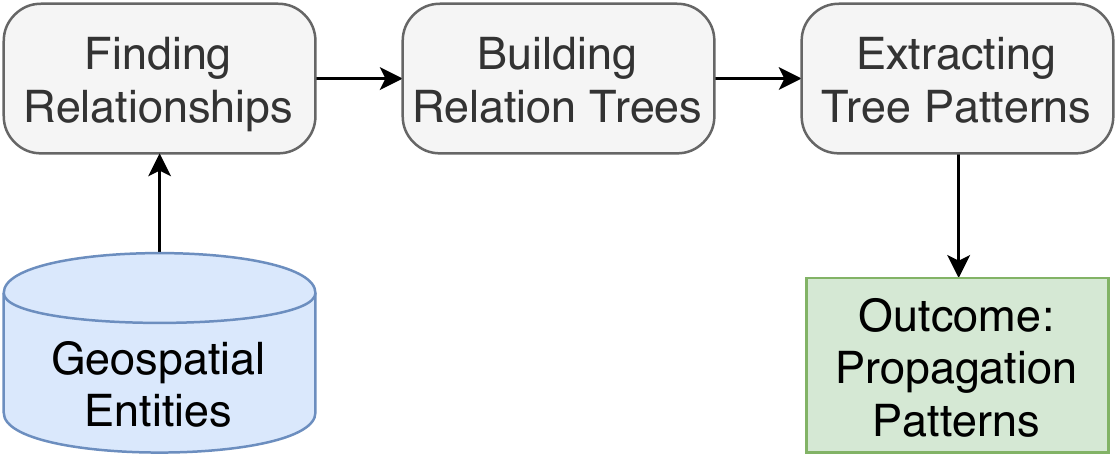} \vspace{-3pt}
        \caption{\small Short-term pattern discovery}
        \label{fig:short_term}
    \end{subfigure}\hspace{-5pt}
    \begin{subfigure}[b]{0.33\textwidth}
        \includegraphics[scale=0.46]{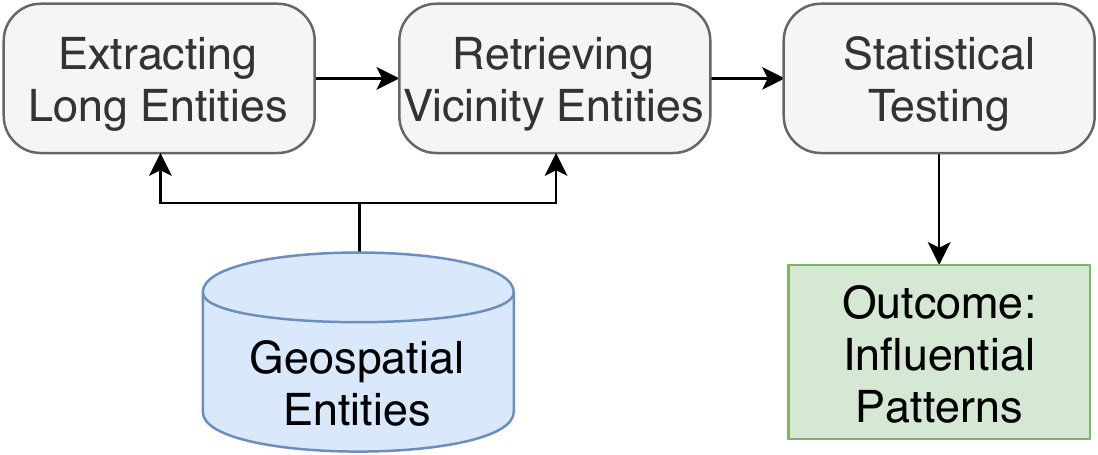} \vspace{-3pt}
        \caption{\small Long-term pattern discovery}
        \label{fig:long_term_process}
    \end{subfigure}\hspace{-5pt}
    \begin{subfigure}[b]{0.33\textwidth}
        \hspace{-2pt}
        \includegraphics[scale=0.44]{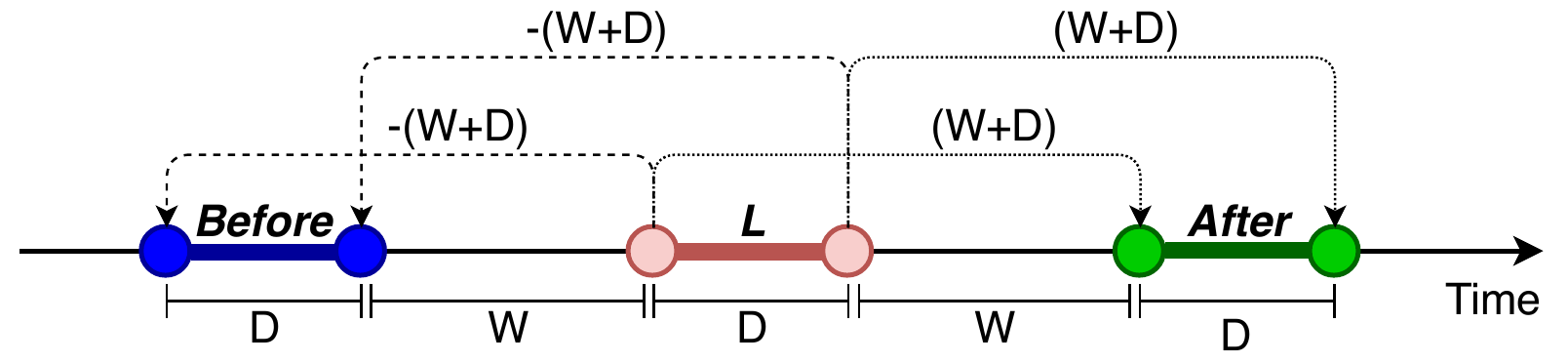}\vspace{-3pt}
        \caption{\small Definition of {\em before} and {\em after} time intervals with respect to long entity {\em L}. ``D'' and ``W`` are short for Day and Week, respectively}
        \label{fig:time_before_after}
    \end{subfigure} \vspace{-5pt}
    \caption{\small Pattern discovery processes for geo-spatiotemporal data (a) and (b); Defining ``before'' and ``after'' time intervals (c).}
    \vspace{-5pt}
\end{figure*}

\section{Pattern Discovery Framework}
\label{sec:method}
In this section we describe the pattern discovery framework, which consists of two major parts, one for discovery of propagation patterns and the other for influential patterns\footnote{All the implementations in Python are available on GitHub: \url{https://github.com/sobhan-moosavi/ShortLongTerm}.}.

\vspace{-5pt}
\subsection{Short-Term Pattern Discovery}
\label{sec:short}
We employed a multi-step process to discover short-term (propagation) patterns in geo-spatiotemporal data. Figure~\ref{fig:short_term} illustrates the process, which includes: 1) finding {\em child-parent} relationships; 2) building {\em relation trees}, and 3) extracting {\em frequent tree patterns}. 

\begin{itemize}[leftmargin=*]
    \item \textbf{Finding Child-Parent Relationships}: The first step is to extract all the weakly dependent pairs of entities to define the child-parent relationship for each pair, using the definitions in Section~\ref{sec:prob}. 
    \item \textbf{Building Relation Trees}: The next step is to create relation trees from the extracted child-parent relations. Here, tree is a rooted, labeled, un-ordered tree (see Section~\ref{sec:prob}). This step results in a forest of relation trees. 
    \item \textbf{Extracting Frequent Tree Patterns}: The last step is to perform frequent tree pattern mining. As described in Section~\ref{sec:prob}, the goal is to extract all frequently observed un-ordered subtrees in our database of relation trees. Examples of such tree patterns with minimum support $75\%$ are shown in Figure~\ref{fig:forest}-b. Here we adopt the SLEUTH algorithm, a growth-based approach proposed by Zaki \cite{zaki2005efficiently}, to extract frequent, embedded, un-ordered sub-trees in our database of relation trees. %The main reasons behind using this particular algorithm are: 1) it is proved to extract all the frequent embedded un-ordered tree patterns in a dataset \cite{zaki2005efficiently}; and 2) the speed of the algorithm is sufficiently fast even for a large set of input trees, with the least memory usage. 
\end{itemize}

\vspace{-5pt}
\subsection{Long-Term Pattern Discovery}
\label{sec:long}
In this section we describe the process of long pattern discovery to study the magnitude of impact of long-term entities. This process (shown in Figure~\ref{fig:long_term_process}) consists of three steps; 1) extracting long-term entities, 2) retrieving vicinity entities, and 3) performing statistical significance testing to explore influential patterns.   

\begin{itemize}[leftmargin=*]
    \item \textbf{Extracting Long Entities}: A long (or long-term) entity is one that last for a long time interval, defined by a heuristic threshold. To define such threshold, we first obtain the distribution of duration of entities over the input dataset; and then consider the $99^{th}$ percentile of the distribution as the threshold which defines long entities. Next, we resolve time and spatial {\em overlaps} between long geospatial entities using Algorithm~\ref{algo:merge_overlaps}, to identify and merge overlaps. In this algorithm, we first identify all the conflicted cases for an entity $l$ (lines 2--7); then merge the conflicted entities by updating time, location, and type of $l$ (lines 8--11); and finally we update the list of long entities (line 12). Function $co\text{-}occurrence(.)$ checks the time-overlap between two entities, and function $collocation(.)$ checks the geographical collocation, using distance threshold $\rho$. 
    
    \item \textbf{Retrieving Vicinity Entities}: After extracting long entities and resolving overlaps, we retrieve entities in the vicinity of each long entity. Thus, given a long entity $L$, we need to find subsets $\mathcal{S}_R$, $\mathcal{S}\textit{--before}_R$, and $\mathcal{S}\textit{--after}_R$ as follows ($R$ is a maximum vicinity distance): 
    \begin{itemize}[leftmargin=5pt]
        \item [\textbf{--}] $\mathcal{S}_R$: for this set we look for all those geospatial entities which happened within a distance $R$ from $L$, with start time strictly after the start time of $L$, and finished before the end time of $L$. 
        
        \item [\textbf{--}] $\mathcal{S}\textit{--before}_R$: this set is similar to the previous one, except we pick a different time interval to define vicinity, as shown in Figure~\ref{fig:time_before_after}. Based on this process, we move start and end time of $L$ to $W+D$ days before, where $W$ stands for one week, and $D$ shows duration of $L$ in days. In such an interval, we extract all the entities which happened in vicinity distance $R$ from $L$.
        
        \item [\textbf{--}] $\mathcal{S}\textit{--after}_R$: similar to the previous one, except we move the start and end time of $L$ to $W+D$ days after. 
    \end{itemize}
    
    \item \textbf{Mining Patterns by Statistical Testing}: Given the set of long entities, we first categorize them into disjoint buckets based on a common characteristic or criteria (e.g., their location or their type). Then, for each bucket, we compare the values of $\mathcal{S}_R$, $\mathcal{S}\textit{--before}_R$, and $\mathcal{S}\textit{--after}_R$ for all long entities, to determine whether there is any significant difference, therefore impact. For this purpose, we design six different testing scenarios and use {\em two-sample t-test} to test the difference between sample means. For a bucket $B$, we first calculate the following mean values: $\mu_L$, $\mu_{before}$, and $\mu_{after}$ as average of $\mathcal{S}_R$, $\mathcal{S}\textit{--before}_R$, $\mathcal{S}\textit{--after}_R$, respectively, based of the long entities in bucket $B$. Further, we take the average of $\mathcal{S}\textit{--before}_R$ and $\mathcal{S}\textit{--after}_R$ for each long entity in bucket $B$, and take the average of average values and denote that by $\mu_{avg}$. Now, we define the following tests for bucket $B$: 
    \begin{itemize}[leftmargin=5pt]
        \item [\textbf{--}] \textbf{$T_1$}: $\mu_{avg} = \mu_L$ versus $\mu_{avg} < \mu_L$. A one-sided test which examines whether the number of geospatial entities during a long entity is larger than this number when there is not such a long entity. 
        
        \item [\textbf{--}] \textbf{$T_2$}: $\mu_{avg} = \mu_L$ versus $\mu_{avg} > \mu_L$. Similar to the previous one, but with the opposite alternative hypothesis. 
        
        \item [\textbf{--}] \textbf{$T_3$}: $\mu_{before} = \mu_L$ versus $\mu_{before} < \mu_L$. A one-sided test which examines whether the number of geospatial entities during a long entity is larger than when the long entity is not started yet. 
        
        \item [\textbf{--}] \textbf{$T_4$}: $\mu_{before} = \mu_L$ versus $\mu_{before} > \mu_L$. Similar to the previous one but with the opposite alternative hypothesis.
        
        \item [\textbf{--}] \textbf{$T_5$}: $\mu_{after} = \mu_L$ versus $\mu_{after} < \mu_L$. A one-sided test which examines whether the number of geospatial entities during a long entity is larger than when the long entity is ended. 
        
        \item [\textbf{--}] \textbf{$T_6$}: $\mu_{after} = \mu_L$ versus $\mu_{after} > \mu_L$. Similar to the previous one but with the opposite alternative hypothesis. 
    \end{itemize}
\end{itemize}

\noindent Note that in all of the above tests, the first condition is the {\em null hypothesis} and the second one is the {\em alternative hypothesis}. 

\begin{algorithm}[ht]
    \small
    \DontPrintSemicolon
    \KwIn{Long entity set $\mathcal{L}$, and distance threshold $\rho$}
    \For {$l$  \textbf{in}  $\mathcal{L}$}
    {  
        $List = []$\\
        \For {l' \textbf{in}  $\mathcal{L}$}
        {
            \If {$\text{co-occurrence}(l,l')$ \textbf{and} $\text{collocation}(l,l',\rho)$}
                %~~~$\big(l.airport == l'.airport$ \textbf{ or }$distance(l, l') \leq$ $\rho$\big)}
                {
                    List.add(l')
                }
        }
        $l.StartTime  = \min_{\forall e \in List} StartTime(e)$
        
        $l.EndTime    = \max_{\forall e \in List} EndTime(e)$
        
        $l.location   = center_{\forall e \in List}(List)$
        
        $l.Type       = concat_{\forall e \in List}(e.Type) $
        
        $\mathcal{L} = \mathcal{L}-List$
    }
    \KwOut{$\mathcal{L}$}
    \caption{Merge Geospatial Overlaps}
    \label{algo:merge_overlaps}
\end{algorithm}

\vspace{-5pt}
\section{Experiments and Results}
\label{sec:results}
In this section, we describe how the proposed framework was employed to perform pattern discovery. We start with the short-term pattern discovery, and then describe the results for the long-term pattern discovery. 

\begin{figure*}[ht!]
    \vspace{-5pt}
    \small
    \centering
    \setlength\tabcolsep{2pt}
    \begin{tabular}{c  c  c | c  c  c | c c c | c c c}
        \begin{tabular}{@{}c@{}} \textbf{Pattern: 1} \\ \#States: 49 \\ Avg. Sup: 52\% \\ Peak-time: 15-17\\ Road: Mixture\end{tabular} & \raisebox{-.5\height}{\includegraphics[scale=0.3]{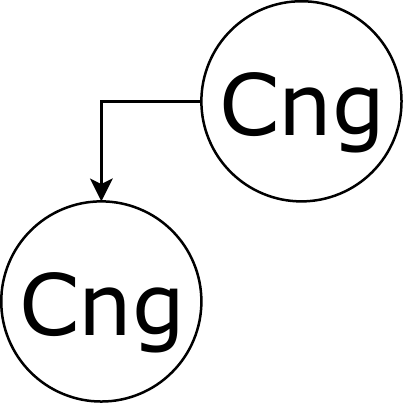}} &&
        \begin{tabular}{@{}c@{}} \textbf{Pattern: 2} \\ \#States: 49 \\ Avg. Sup: 23\% \\ Peak-time: 15-17\\ Road: Mixture\end{tabular} & \raisebox{-.5\height}{\includegraphics[scale=0.27]{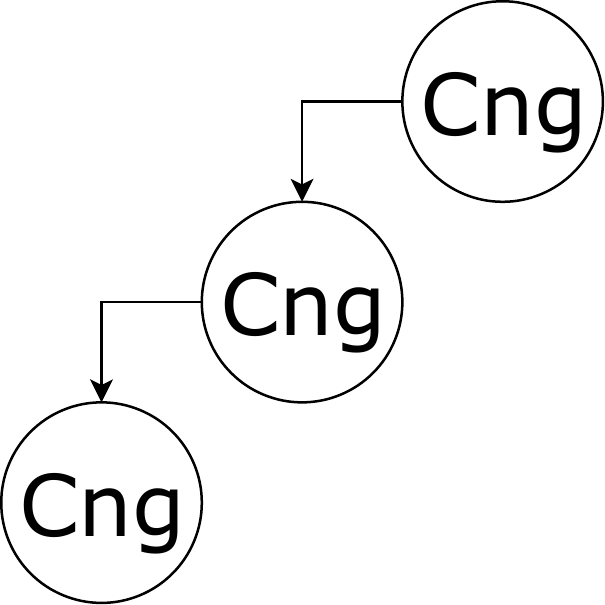}} &&
        \begin{tabular}{@{}c@{}} \textbf{Pattern: 3} \\ \#States: 48 \\ Avg. Sup: 26\% \\ Peak-time: 13-17\\ Road: Mixture\end{tabular} & \raisebox{-.5\height}{\includegraphics[scale=0.3]{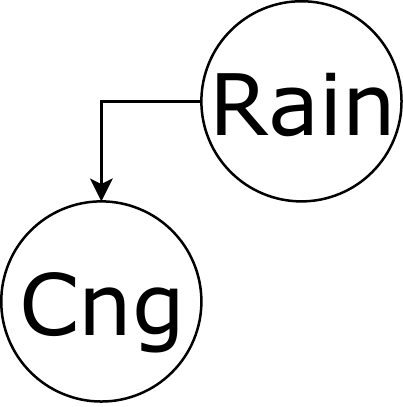}} &&
        \begin{tabular}{@{}c@{}} \textbf{Pattern: 4} \\ \#States: 47 \\ Avg. Sup: 21\% \\ Peak-time: 15-17\\ Road: Mixture\end{tabular} & \raisebox{-.5\height}{\includegraphics[scale=0.3]{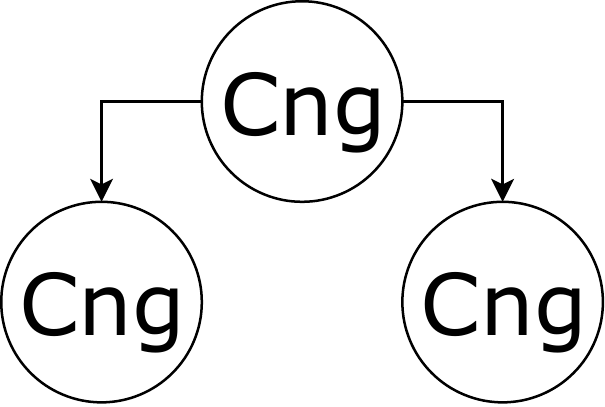}} \\
        \hline
        \hline
        \begin{tabular}{@{}c@{}} \textbf{Pattern: 5} \\ \#States: 37 \\ Avg. Sup: 32\% \\ Peak-time: 5-9\\ Road: Interstates \\and Freeways\end{tabular} & \raisebox{-.5\height}{\includegraphics[scale=0.3]{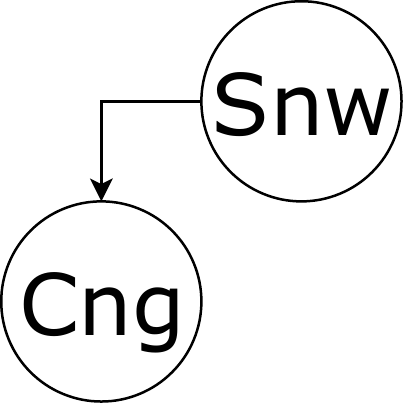}} &&
        \begin{tabular}{@{}c@{}} \textbf{Pattern: 6} \\ \#States: 31 \\ Avg. Sup: 24\% \\ Peak-time: 5-9\\ Road: Interstates \\and Freeways\end{tabular} & \raisebox{-.5\height}{\includegraphics[scale=0.3]{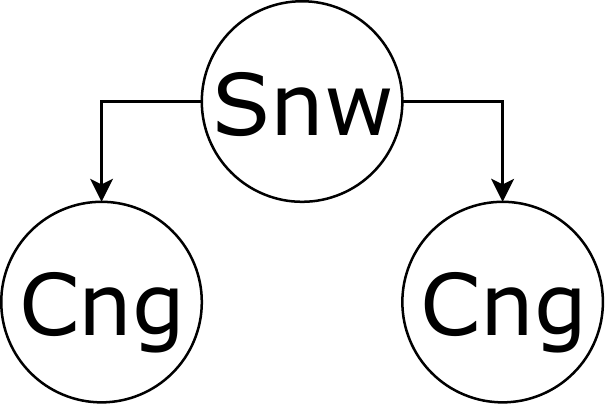}} &&
        \begin{tabular}{@{}c@{}} \textbf{Pattern: 7} \\ \#States: 30 \\ Avg. Sup: 24\% \\ Peak-time: 6-9\\ Road: Mixture\end{tabular} & \raisebox{-.5\height}{\includegraphics[scale=0.3]{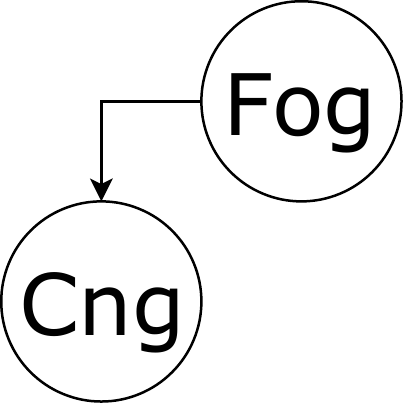}} &&
        \begin{tabular}{@{}c@{}} \textbf{Pattern: 8} \\ \#States: 27 \\ Avg. Sup: 16\% \\ Peak-time: 15-17\\ Road: Cities\end{tabular} & \raisebox{-.5\height}{\includegraphics[scale=0.27]{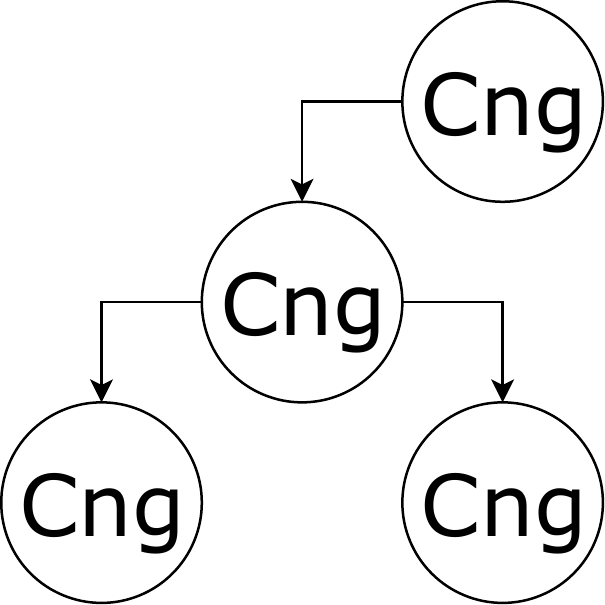}} \\
        \hline
        \hline
        \begin{tabular}{@{}c@{}} \textbf{Pattern: 9} \\ \#States: 26 \\ Avg. Sup: 10\% \\ Peak-time: 15-17\\ Road: Cities\end{tabular} & \raisebox{-.5\height}{\includegraphics[scale=0.3]{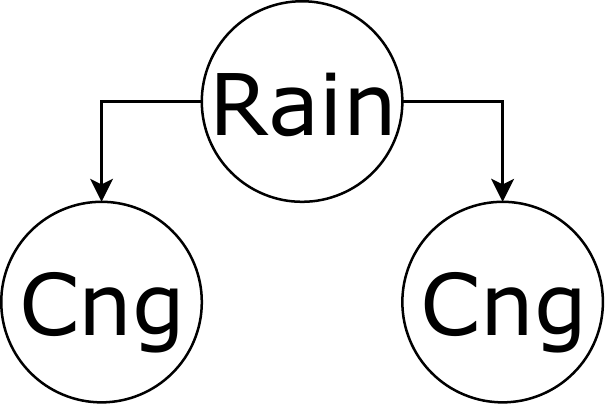}} &&
        \begin{tabular}{@{}c@{}} \textbf{Pattern: 10} \\ \#States: 26 \\ Avg. Sup: 16\% \\ Peak-time: 15-17\\ Road: Cities\end{tabular} & \raisebox{-.5\height}{\includegraphics[scale=0.27]{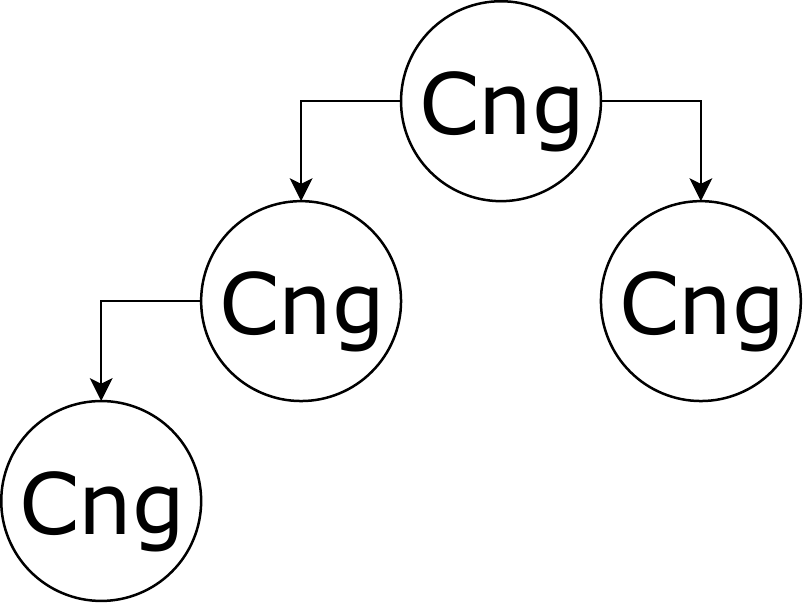}} &&
        \begin{tabular}{@{}c@{}} \textbf{Pattern: 11} \\ \#States: 23 \\ Avg. Sup: 15\% \\ Peak-time: 15-17\\ Road: Cities\end{tabular} & \raisebox{-.5\height}{\includegraphics[scale=0.27]{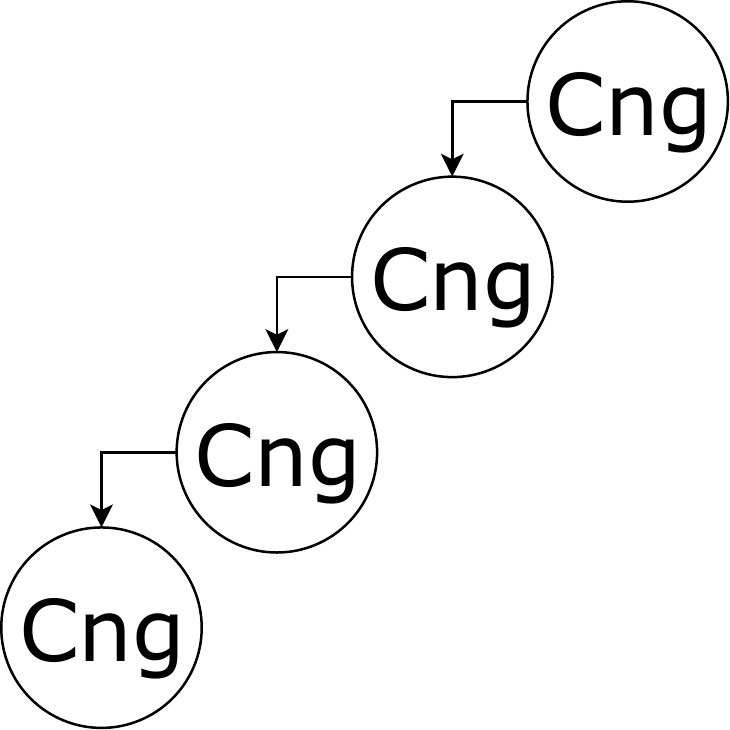}} &&
        \begin{tabular}{@{}c@{}} \textbf{Pattern: 12} \\ \#States: 23 \\ Avg. Sup: 19\% \\ Peak-time: 6-9\\ Road: Cities\end{tabular} & \raisebox{-.5\height}{\includegraphics[scale=0.3]{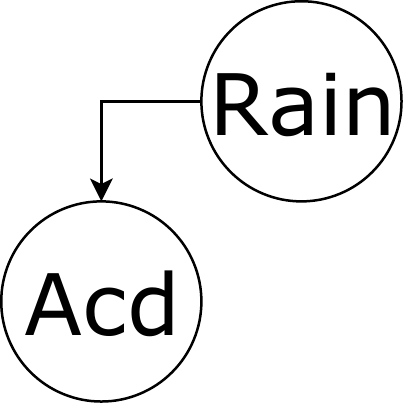}} \\ 
        %&&&&&&&\\
    \end{tabular}
    \caption{\small Top frequent embedded tree patterns found based on the short-term dependency relation trees. ``Cng'', ``Acd'', and ``Snw'' are short for congestion, accident, and snow. These patterns show the propagation of traffic/weather entities on a short-term basis.}
    \label{fig:short_term_top}
    \vspace{-5pt}
\end{figure*}

\vspace{-5pt}
\subsection{Short-term Pattern Discovery Results}
First, we extracted all short-term child-parent relationships using thresholds $D\text{-}thresh = 300$ {\em meters} and ${T\text{-}thresh = 10}$ {\em minutes} (see Section~\ref{sec:prob})\footnote{For entity type {\em snow}, we empirically set ${T\text{-}thresh = 40}$ minutes, because we expect to see a longer impact of snow on traffic flow.}. These thresholds were found empirically, with $D\text{-}thresh$ ensuring spatial closeness, and $T\text{-}thresh$ is large enough to consider the delay in a ``cause and effect'' type of relationship, with respect to our application domain. Using these settings, we found $5,952,729$ Child-Parent relationships from $15,325,659$ traffic and weather entities. In total, $39.33\%$ of the traffic entities were found to have at least one weakly dependent weather or traffic entity, and $12.82\%$ of weather entities had at least one weakly dependent traffic entity. 
Next, we created $1,723,637$ trees out of $5,952,729$ child-parent relations. The maximum number of nodes in a tree was found to be $25$. Where a traffic entity $t$ had more than one parent, we randomly picked one of them. Given the size of the data and the number of trees, we do not believe that any existing frequent pattern would be missed by this choice. Finally, we employed SLEUTH (Zaki \cite{zaki2005efficiently}) to extract frequent tree patterns at the {\em city-level}. More scalable alternatives \cite{tatikonda2006trips,tatikonda09} were an option but not required for our purpose. After extracting frequent patterns for a city, we used these patterns as {\it core} frequent patterns of the corresponding state. This allows us to account for the potential diversity among different cities in a state (i.e., based on population, traffic, and/or weather condition). As an alternative, if we had chosen core patterns using state as the granularity level, the framework may not identify those patterns which are frequent in one city but infrequent in the others. To choose the minimum-support value, regarding the {\em large size} of data and potential {\em seasonality} in observations, we followed the approach proposed by Fournier-Viger \cite{fournier2010modele}. Based on this approach, we used Equation~\ref{eq:min_sup} to find the minimum support, where $a$, $b$, and $c$ are the positive constants which we empirically set to $0.004$, $1.5$, and $0.05$, respectively. In this formula, $x$ is the number of relation trees in a set, and the minimum relative support is $5\%$.
\begin{equation}
    \label{eq:min_sup}
    \text{min\_sup} = e^{-(ax + b)} + c
\end{equation}
Using the above setting, we extracted $708$ frequent tree patterns for the contiguous United States. In total, there were $90$ unique frequent patterns, with the minimum number of nodes in a tree pattern being 2 and the maximum being 7. Figure~\ref{fig:short_term_top} shows the top 12 frequent tree patterns\footnote{Check \url{https://bit.ly/2Ef8tu7} for the list of all short-term frequent patterns in our data.}. Along with each pattern is shown the number of states which have occurrences of that pattern, the average support value, the peak time for instances of the pattern, and type of the road-network in which instances of a pattern were common. A road-network can be a road inside a city (cities), an interstate or freeway which connects different cities or states to each other, or a mixture of both. Each pattern shows how short-term entities are propagated in a region. For instance, pattern 1 shows a congestion which caused another congestion, and pattern 2 shows a propagation pattern of a chain of traffic congestion entities. % in adjacent locations. 

In total, 50 of 90 unique frequent patterns were initiated by a weather event, where 17 of these patterns were initiated by rain, 14 by snow, 11 by fog, and 8 by the other types of weather entities. These observations demonstrate the significant impact of weather on traffic. While this has been frequently discussed in prior research \cite{brijs2008studying,cools2010assessing,ding2017detecting}, in our work we reveal the propagation patterns which show HOW these weather entities impact traffic. For example, snow-initiated patterns usually happen on interstates and freeways, while rain-initiated patterns happen within roadways inside cities. Also, most complex congestion-related patterns happen within cities road-network, with the average support of patterns which happen in a city is lower than those which happen on interstates and freeways, or the entire road-network. The peak time for the majority of congestion-related patterns was the afternoon rush hour. For weather initiated patterns (except for the rain-related cases), the peak time was the morning rush hour. It was interesting to note that some weather events caused more traffic issues in the morning rather than the afternoon. %These are interesting observations where a weather event usually causes more traffic issues on morning rather than afternoon.

To further analyze the short-term patterns, we created a one-hot vector of size 90 for each state which represents the presence or absence of each unique short-term pattern. By applying K-means clustering \cite{han2011data} on these vectors, we categorized different states based on their short-term propagation patterns. To find the best number of clusters, we adapted Description Length (DL) for K-Means \cite{mdl_kmeans}, which is represented by Equation~\ref{eq:mdl}. In this equation, $p(.)$ is the probability density function based on distance of each data point $x$ from its cluster center $c_x$; $P$ is the number of parameters of distribution function; $K$ is the number of clusters; and $X$ is the set of all data points. By assuming the distribution function for distance from cluster centers to be a Gaussian distribution, we have $P=2$. By choosing $K$ from set $[2, 3, \dots, 10]$, we found the optimal number of clusters to be $4$, which provides the minimum description length. 
\begin{equation}
    \label{eq:mdl}
    DL(K) = -\sum_{x \in X} log(p(||x - c_x||)) + \frac{1}{2}P\text{ }log(|X|) + K\text{ }log(|X|)
\end{equation}

\begin{table}[h]
    \vspace{-3pt}
    \small
    \centering
    \caption{\small Clustering of 49 states into 4 clusters based on their short-term patterns, using K-Means.}\vspace{-8pt}
    \setlength\tabcolsep{12pt} % default value: 6pt
    \begin{tabular}{| c | c |}
        \hline
        \rowcolor{Gray}
        \textbf{Cluster} & \textbf{States} \\
        \hline
        Cluster 1 & \begin{tabular}{@{}c@{}} AL, AR, CT, DC, DE, IA, IN, LA, MA, ME, MI, MN,  \\ MO, MS, NE, NH, OH, OK, RI, SD, TN, VT, WI \end{tabular}\\
        \hline
        Cluster 2 & \begin{tabular}{@{}c@{}} AZ, CO, ID, KS, KY, MD, MT, NC, ND, NJ, \\ NM, NV, OR, PA, SC, UT, VA, WV, WY \end{tabular} \\
        \hline
        Cluster 3 & FL, GA, IL, NY, TX, WA\\
        \hline
        Cluster 4 & CA\\
        \hline
    \end{tabular}
    \label{tab:state_clustering}
    \vspace{-3pt}
\end{table}

\noindent Table~\ref{tab:state_clustering} shows the result of clustering, in which we profile clusters as follows:
\begin{itemize}[leftmargin=*]
    \item \textbf{Cluster 1}: mostly contains states with fewer traffic incidents (as related to weather). These are either states with lower population (e.g., NE, SD, etc.); or states where the impact of weather is mitigated by effective road crews (e.g., OH, MN, etc.).
    \item \textbf{Cluster 2}: mostly contains states with considerably more traffic issues in comparison to the states in cluster 1. Distinguished patterns which only observed for this cluster are chain of accidents, and complex snow-initiated patterns. 
    \item \textbf{Cluster 3}: contains states with at-least one major city with significant traffic issues. Distinguished patterns observed for this cluster are those which initiated by construction, rain, severe-cold, and storm.  
    \item \textbf{Cluster 4}: contains only one state whose traffic patterns bore no similarity to any other state. Majority of distinguished patterns of this cluster are complex congestion-related, fog-initiated, and flow-incident related ones. 
\end{itemize}

\noindent It is worth noting that the states which were clustered together were not necessarily located in the same geographical region, and might not have the same weather condition during the different seasons. However, their propagation patterns of traffic and impact of weather on traffic was found to be the same, which led to them being in the same cluster. 
\vspace{-5pt}
\subsection{Long-term Pattern Discovery Results}
\subsubsection{Parameter Settings and Conventions}
As described in Section~\ref{sec:long}, first we use the $99^{th}$ percentile of distribution of the duration of entities across the entire dataset, as the threshold to extract long entities -- resulting in about $300$ minutes. Using this threshold, we extracted $280,649$ long entities. To merge the overlaps by Algorithm~\ref{algo:merge_overlaps}, we set $\rho = R$, where $R$ is the spatial neighborhood distance to define sets $\mathcal{S}_R$, $\mathcal{S}\textit{--before}_R$, and $\mathcal{S}\textit{--after}_R$. In this way, we ensure that after the merge, there is no pair of long entities whose spatial neighborhood overlapped. Next we describe how to determine $R$, and then perform merging the overlaps. 

\vspace{5pt}
\noindent\textbf{Extracting $R$ for long Traffic entities.} 
To determine $R$, we use a random sample $S_1$ of two million traffic entities, and apply DBSCAN \cite{han2011data} to cluster entities in set $S_1$. We find the radius of each cluster as the maximum distance from the center, and obtain $R$ as the average radius across all clusters. To define the two DBSCAN parameters -- $\epsilon$ (maximum neighborhood distance) and {\em minPts} (the minimum required number of neighbors for not being an outlier), we use Algorithm~\ref{alg:dbscan_parameters} adapted from \cite{spark_dbscan}. Using a random sample set of $0.5$ million traffic entities in terms of $S_2$, we obtained $\epsilon = 4.09$ $\textit{miles}$ and $minPts = 463$. Applying DBSCAN on $S_1$ resulted in 191 clusters, with the average radius $R$ of these clusters being $14.03$  $\textit{miles}$. 

\begin{algorithm}[ht]
    \small
    \caption{Finding DBSCAN Parameters}
    \begin{algorithmic}[1]
        \STATE Input: $S_2$, a large sample of traffic entities. 
        \STATE In $S_2$, obtain the closest neighbor distance for each entity, and let $\mathrm{C}_1$ be the $99^{th}$ percentile of distribution of the closest neighbor distances. 
        \STATE For each entity, count the number of entities within distance $\mathrm{C}_1$, and obtain distribution of count values over $S_2$. 
        \STATE Let $\mathrm{C}_2$ be the $99^{th}$ percentile of distribution of the count values. 
        \STATE Output: $\mathrm{C}_1$ as $\epsilon$, and $\mathrm{C}_2$ as $minPts$
    \end{algorithmic}
    \label{alg:dbscan_parameters}
\end{algorithm}

\vspace{5pt}
\noindent Note that we cannot quantitatively define $R$ for long weather entities. Thus, we define a traffic entity $t$ be within $R\textit{--neighborhood}$ of a long-term weather entity $w$, if $t$'s zipcode can be mapped to the airport station which $w$ is reported from, as the closest station. With $\rho = 14.03$, and after merging the overlaps, we ended up with $148,237$ long entities. Table~\ref{tab:long_entities_stats} provides the details on top-15 types of long entities, before and after the merge. Note that after the merge process, some of the types were combined to generate new type labels (e.g., Rain\_Event). Next, setting $R=14.03$, we created vicinity sets $\mathcal{S}_R$, $\mathcal{S}\textit{--before}_R$, and $\mathcal{S}\textit{--after}_R$ for each long-term entity. 

\vspace{5pt}
\noindent \textbf{Bucketing.} Prior to employing statistical significance testing to identify long-term patterns, we need to determine the buckets of long-term entities. We use three different criteria to create disjoint buckets, namely, {\em Location}, {\em Duration}, and {\em Type}. Each ``Location'' bucket contains all the long entities which occurred in the same state. For the ``Duration bucket'', we first define several duration buckets (intervals), and then assign each long entity to a bucket. For ``Type buckets'', we create buckets of long entities, where each bucket contains all the entities of the same type. 

\vspace{5pt}
\noindent \textbf{Positive and Negative Impacts.} The positive (negative) impact refers to the case where the value of $\mathcal{S_R}$ is larger (lower) than $\mathcal{S}\textit{--before}_R$, $\mathcal{S}\textit{--after}_R$, or their average. A significant positive impact can be determined by tests $T_1$, $T_3$, or $T_5$. A significant negative impact can be determined by tests $T_2$, $T_4$, or $T_6$.

\begin{table}[h]
    %\small
    \centering
    \setlength\tabcolsep{9pt} % default value: 6pt
    \caption{\small Top 15 long entity types and their frequency.} \vspace{-8pt}
    \scalebox{0.8}{
        \begin{tabular}{| c | c | c | c |}
        \hline
        \multicolumn{2}{|c|}{\textbf{Before Merge}}   & \multicolumn{2}{|c|}{\textbf{After Merge}}\\
        \hline
        \hline
        Type   &    Frequency  & Type   &    Frequency    \\
        \hline
        Construction         &  113,984 & Rain												 &  38,253 \\
        Rain                 &   41,668 & Snow                                               &  25,820 \\
        Event                &   32,144 & Construction                                       &  22,373 \\
        Snow                 &   27,723 & Fog                                                &  19,553 \\
        Fog                  &   20,847 & Event                                              &  16,753 \\
        Congestion           &   17,314 & Severe-Cold                                        &  11,671 \\
        Flow-Incident        &   13,099 & Congestion                                         &   3,206 \\
        Severe-Cold          &   12,083 & Flow-Incident                                      &   2,381 \\
        Storm                &     733  & Construction\_Event                                &   1,332 \\
        Other                &     440  & Construction\_rain                                 &    758 \\
        Lane-Blocked         &     253  & Storm                                              &    709 \\
        Accident             &     226  & Congestion\_Flow-Incident                          &    675 \\
        Precipitation        &      72  & Congestion\_Event                                  &    516 \\
        Broken-Vehicle       &      49  & Event\_Rain                                        &    514 \\
		Hail                 &      14  & Congestion\_Construction                           &    359 \\
        \hline
        total                & 280,649  & total                                              &  144,873 \\
        \hline
        \end{tabular}
    }
    \label{tab:long_entities_stats}
    \vspace{-3pt}
\end{table}

\begin{figure*}[ht!]
    \centering
    \includegraphics[scale=0.37]{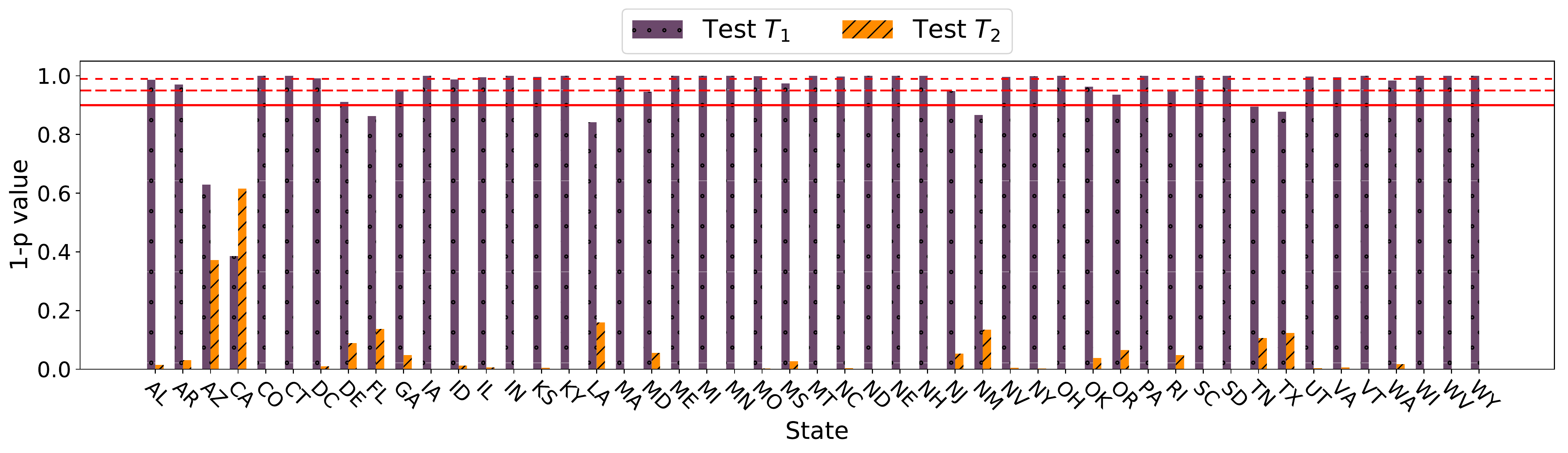}\vspace{-10pt}
    \caption{\small Statistical significance testing by test $T_1$ and $T_2$ for {\em Location} buckets. Red lines show three confidence levels $90\%$, $95\%$, and $99\%$.}
    \label{fig:state}
    \vspace{-8pt}
\end{figure*}

\begin{figure*}
    \minipage{0.33\textwidth}
        \centering
        \includegraphics[width=\linewidth]{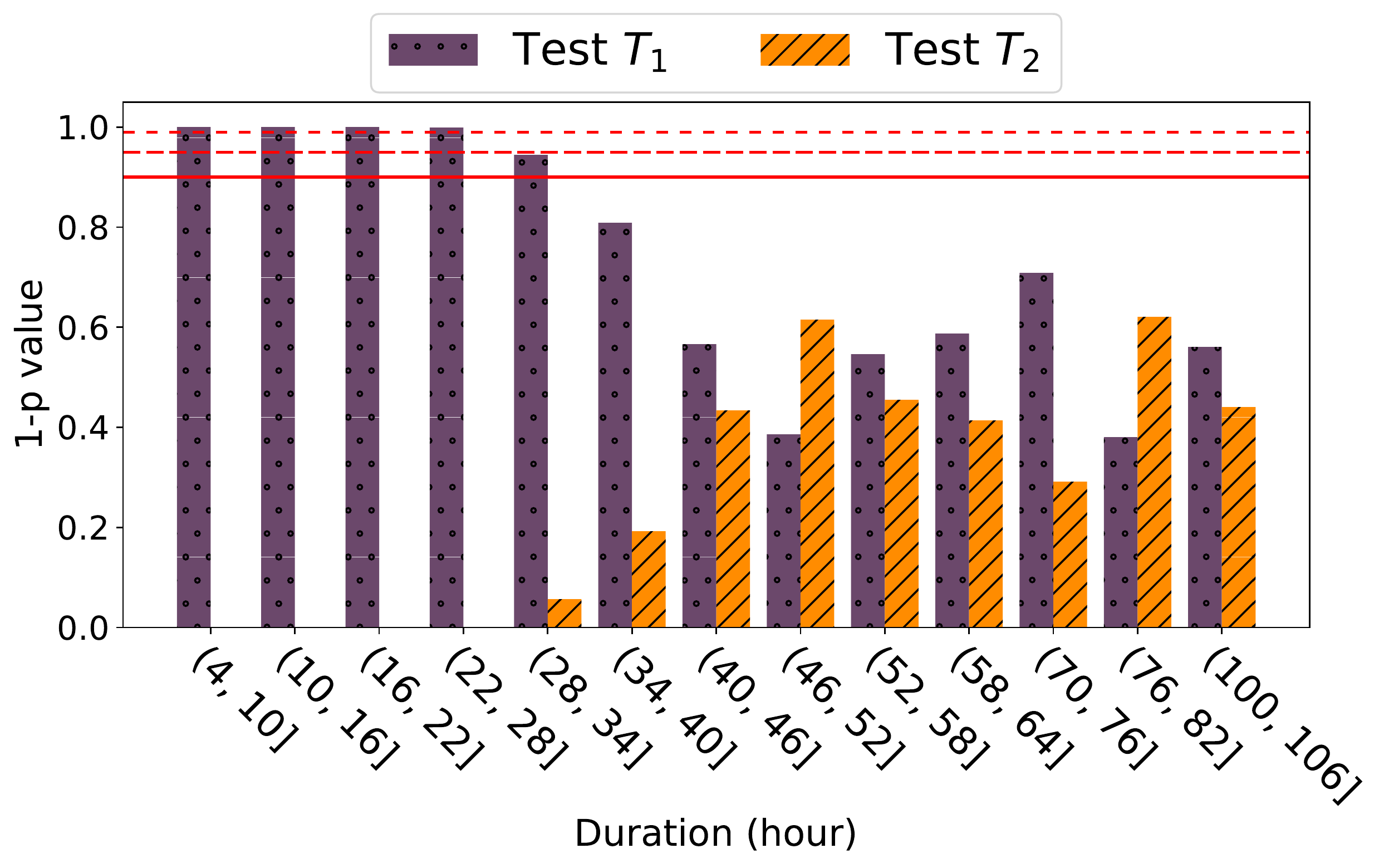}
        (a)
    \endminipage\hfill
    \minipage{0.33\textwidth}
        \centering
        \includegraphics[width=\linewidth]{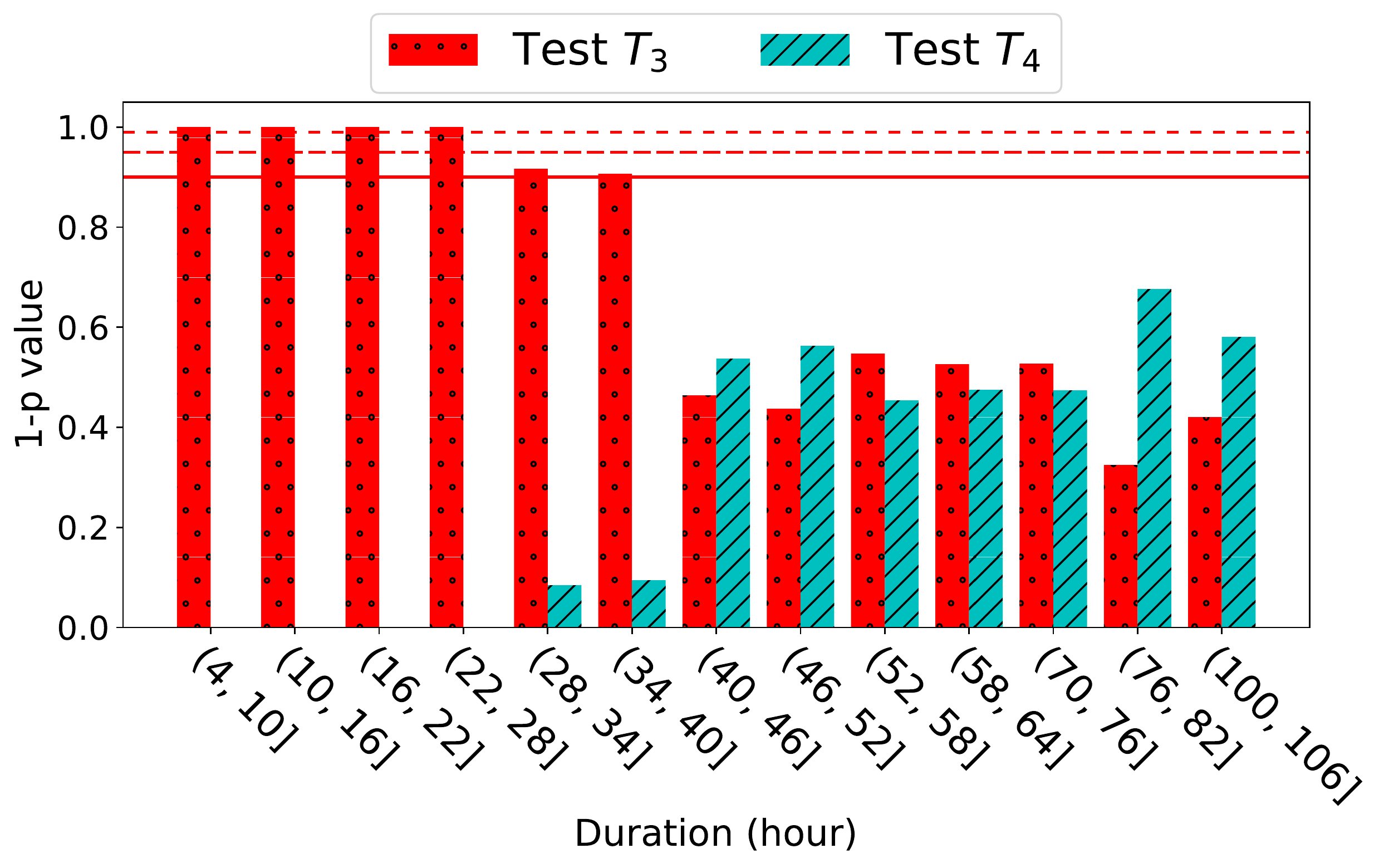}
        (b)
    \endminipage\hfill
    \minipage{0.33\textwidth}
        \centering
        \includegraphics[width=\linewidth]{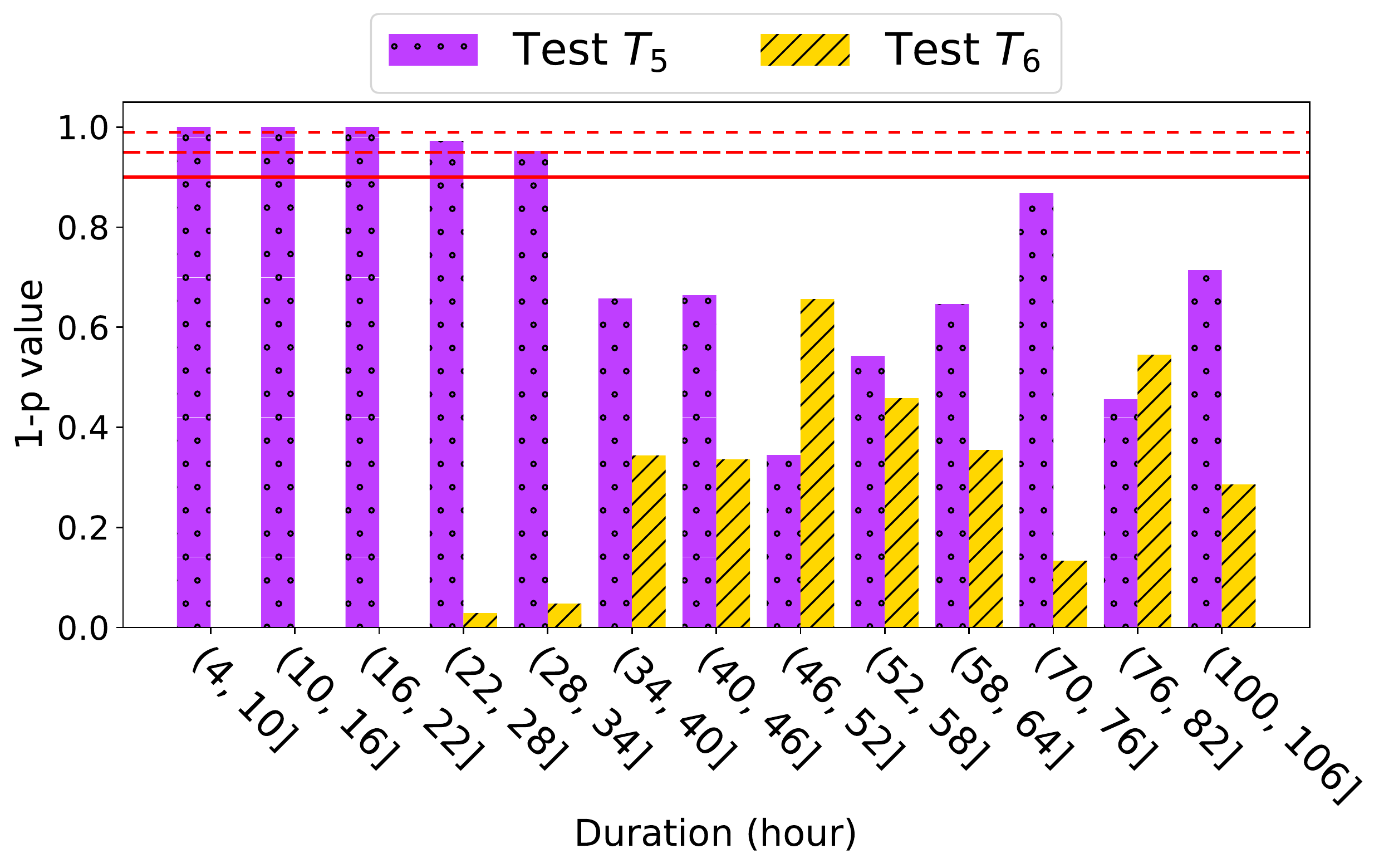}
        (c)
    \endminipage\hfill
    \vspace{-8pt}
    \minipage{\textwidth}
        \centering
        \caption{\small Statistical significance testing for {\em Duration} buckets. Red lines show three confidence levels $90\%$, $95\%$, and $99\%$.}\label{fig:duration}
    \endminipage
    \vspace{-8pt}
\end{figure*}

\begin{figure*}
    \minipage{0.33\textwidth}
        \centering
        \includegraphics[width=\linewidth]{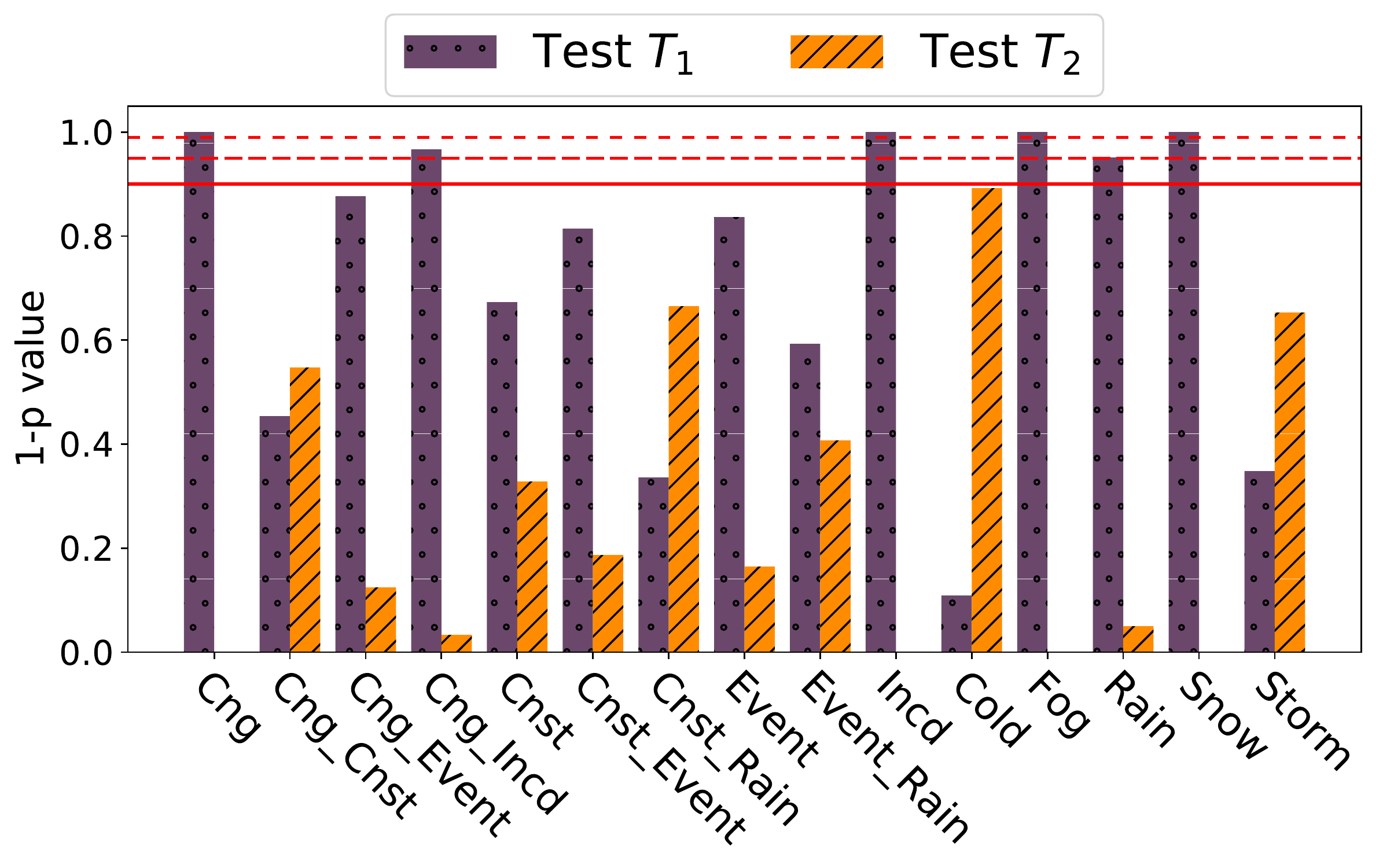}
        (a)
    \endminipage\hfill
    \minipage{0.33\textwidth}
        \centering
        \includegraphics[width=\linewidth]{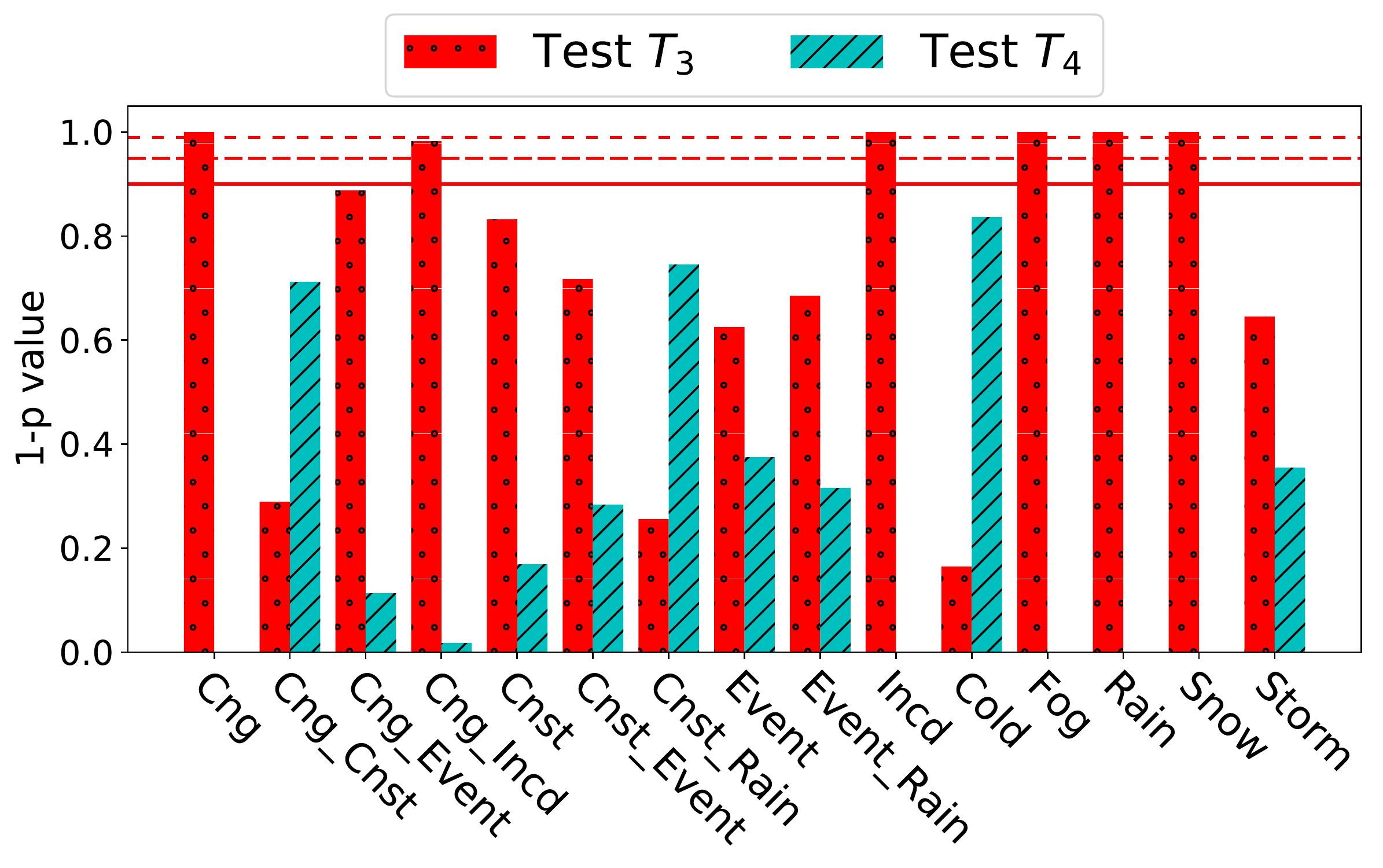}
        (b)
    \endminipage\hfill
    \minipage{0.33\textwidth}
        \centering
        \includegraphics[width=\linewidth]{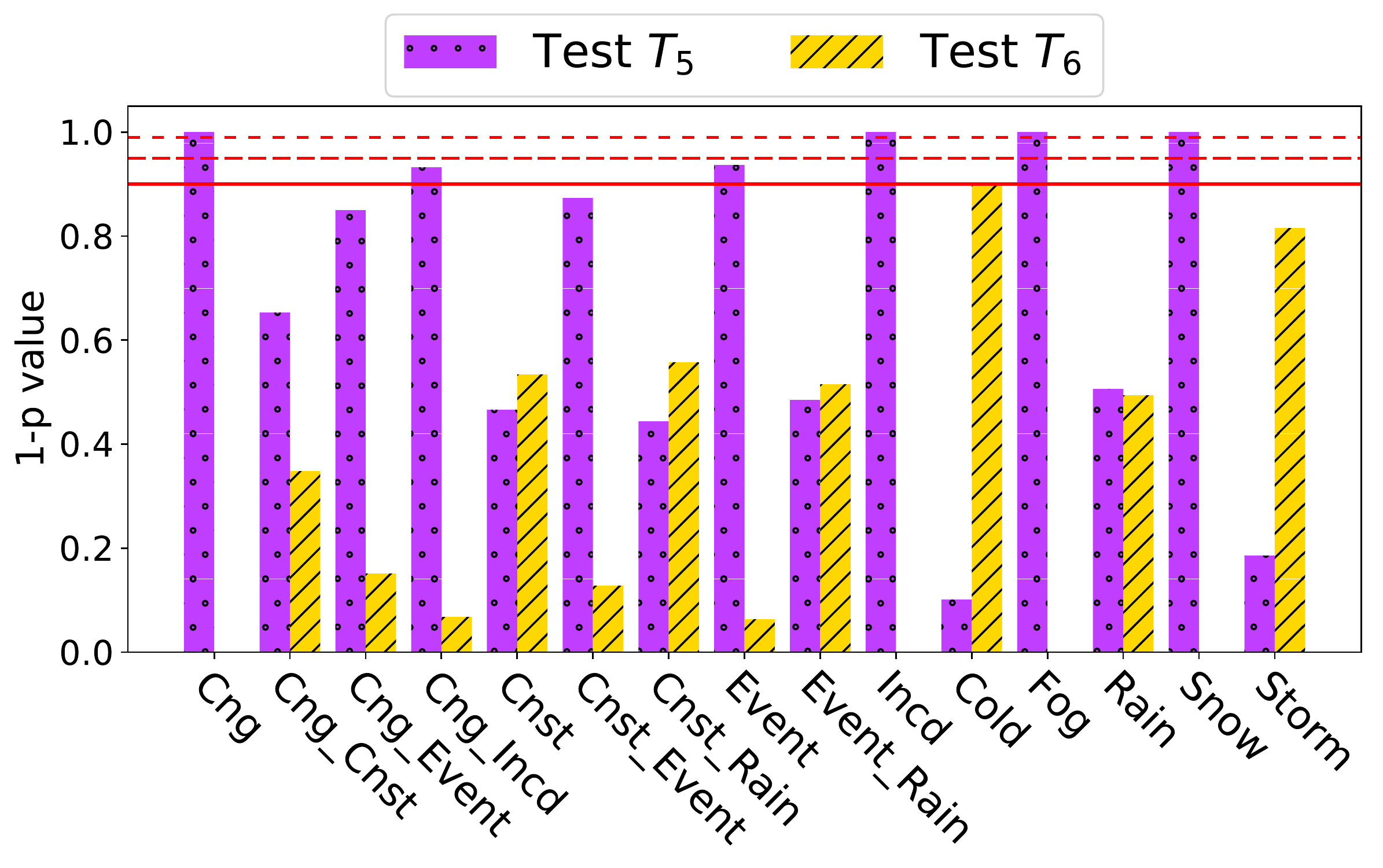}
        (c)
    \endminipage\hfill
    \vspace{-8pt}
    \minipage{\textwidth}
        \centering
        \caption{\small Statistical significance testing for {\em Type} buckets. Red lines show three confidence levels $90\%$, $95\%$, and $99\%$. Cng, Cnst, Incd, and Cold are short for Congestion, Construction, Flow-Incident, and Severe-Cold, respectively.}\label{fig:type}
    \endminipage\hfill
    \vspace{-8pt}
\end{figure*}

\subsubsection{Long-term Patterns}
We present the identified long-term patterns in terms of three categories of such patterns; each category obtained based on a particular bucketing criteria. 

\vspace{5pt}
\noindent {\bf Location-based Patterns.} Using location as the bucketing criteria, we applied significance tests $T_1$ and $T_2$ to identify patterns of the form ``$\textit{long-term entity in location L} \rightarrow \textit{more (or less) traffic incidents}$''. Figure~\ref{fig:state} shows the results of these tests. Here we represent $1-p\text{-}value$, and also show three confidence levels $90\%$, $95\%$, and $99\%$ as three red lines (the results of other tests are not presented because any different trend of results was not observed). For a majority of states, we observed that a long-term weather/traffic entity had a significant impact on traffic flow. In the majority of the cases we found the result of test $T_1$ to be significant, which means a positive impact. Out of the 49 states, we found 30 to be significant with a confidence of $99\%$, 8 with a confidence of $95\%$, and 5 with a confidence of $90\%$. We also found that the existence of long-term traffic or weather entities did not have much impact on traffic flow for AZ, CA, FL, LA, NM, and TX; although three of these states (i.e., CA, FL, and TX) are the top-3 states with the most observed traffic entities. This observation reveals that in a state with more traffic issues, the existence of a long-term incident does not have much impact on traffic flow. Incidentally, CA was the only state for which the p-value is found to be lower by Test $T_2$ (although insignificantly so). This could imply that CA has a unique condition where a long-term weather or traffic entity causes less traffic issues in comparison to the time when there is no such long-term entity. %This last observation is also compatible with our finding in terms of clustering of states based on their short-term patterns (see Table~\ref{tab:state_clustering}). 

\vspace{5pt}
\noindent {\bf Duration-based Patterns.} Using duration of long entities as the bucketing criteria, we applied all the six significance tests to identify patterns of the form ``$\textit{long-term entity with duration D} \rightarrow \textit{more }$ $\textit{(or less) traffic incidents}$''. Figure~\ref{fig:duration} shows the results of these tests. We conclude that the shorter the duration of a long-term entity is, the more significant its impact. Also, for long-term entities which lasted for more than 40 hours, we usually do not observe any significant impact. This observation might be due to adaptation of driving habits to the new conditions. Also, a comparison of the results of tests $T_3$ and $T_4$ with tests $T_5$ and $T_6$, provided evidences of more positive impacts based on the \textit{after} interval, rather than the \textit{before} interval, for long entities which lasted more than 28 hours. Given that a majority of such long entities were construction projects (about $75\%$), we posit two potential interpretations. First, after a long construction project, we tended to observe a smoother traffic flow, even in comparison to the time before the construction event. This observation might be due to the road conditions improving after the construction, but also could point to the fact that, after a long construction project, there might be a significant group of drivers who stuck with the alternative routes discovered when the construction was active. 

\vspace{5pt}
\noindent {\bf Entity-type-based Patterns.} Using type of the long entities as the bucketing criteria, we applied all the six significance tests to identify patterns of the form ``$\textit{long-term entity of type T} \rightarrow \textit{more }$ $\textit{(or less) traffic incidents}$''. Figure~\ref{fig:type} shows the results of these tests. Regarding the weather-based long entities, we observe the significant impact of all available types of weather entities, except for the ``storm'' event. However, we have an interesting diversity among impacts of different types of weather entities. Usually for ``fog'', ``snow'' and ``rain'', based on Tests $T_1$ and $T_2$, we see a positive impact on traffic, while for ``severe-cold'' we observe a negative impact. This observation reveals that in extremely cold temperatures, we should expect to see smoother traffic flow probably because of fewer vehicles on the roads. Tests $T_3$ through $T_6$ also support such conclusion. %Unlike the other impactful types of weather entities, for ``rain'' we only observed positive impact when considered the before time interval (see tests $T_3$ to $T_6$). Although non of the tests provided significant results for ``storm'', we observed the probability of having more traffic issues after the storm is higher, based on tests $T_5$ and $T_6$. 
Regarding the traffic-based long entities, we observed significant impacts by ``congestion'', ``event'', and ``flow-incident''. In case of a long-term ``congestion'', we have positive impact in comparison to before and after. For ``flow-incident'', we also observed a similar situation. However, for a long-term ``event'', we only observed positive impact in comparison to the time when the ``event'' is terminated (test $T_5$). It was interesting to note that a long-term construction had almost no significant impact on traffic flow. However, based on tests $T_3$ and $T_4$, we could expect to see more traffic issues during a long-term construction than before it or after it.  

\section{Conclusion and Future Work}
\label{sec:conc}
To overcome with the shortcomings posed by the existing general-purpose spatiotemporal pattern discovery frameworks, such as relying on a simplistic definition of spatiotemporal neighborhood, we present a new framework to extract {\em propagation} as well as {\em influential} patterns in geo-spatiotemporal data using improved and novel techniques. 
To extract propagation patterns, that indicate immediate impacts, we use a stricter definition of spatial collocation and co-occurrence relationships to create relation trees, and then perform tree pattern mining in a forest of relation trees. Influential patterns, that show lagging impacts, explore the impact of long-lived geospatial entities on their neighborhood, and we used statistical techniques to identify such patterns. Using a new and unique geo-spatiotemporal dataset of traffic and weather entities, which is collected, processed, and augmented for the contiguous United States over two years, we explored 90 prevalent propagation patterns, where 50 of them were initiated by weather (mostly observed in morning) and the rest by traffic entities (mostly observed in afternoon). Based on these patterns, we identified four categories of US states, which show similarity of driving behavior and transportation infrastructures between different states. We also studied the lagging impact of long-term traffic or weather entities, with respect to location, duration, and type of the entities. Interestingly, we identified a positive impact of long-term entities in a majority of the states, except a few ones such as CA, FL, and TX. In general, we found that long-term entities which lasted for at most 40 hours have the maximum impact on traffic flow. We found that long-term congestion, snow, rain, fog, severe-cold, and flow-incidents cause the most significant lagging impact on traffic flow. In terms of future research, we plan to separately study the lagging impact of different entity types for different states and top-cities. 

\section*{Acknowledgment}
This work is supported by a grant from the NSF (EAR-1520870) and
one from the
Ohio Supercomputer Center (PAS0536). Any findings and opinions are those of the authors. We also thank Mr. Abbas Shakiba for the preliminary discussions. 

% placing reference section
\bibliographystyle{ACM-Reference-Format}
\bibliography{main}

\end{document}